\pdfoutput=1

\documentclass[11pt,twoside,a4paper,cmspaper,final,collab]{cms-tdr}
\def\svnVersion{398306}\def\cmsCernNoTag{CERN-EP/2016-225}\def\cmsCernDate{\today}\def\cmsMessage{Published in the Journal of High Energy Physics as \href{http://dx.doi.org/10.1007/JHEP04(2017)018}{\doi{10.1007/JHEP04(2017)018.}}}

\begin{document}\cmsNoteHeader{SUS-14-022}

\hyphenation{had-ron-i-za-tion}
\hyphenation{cal-or-i-me-ter}
\hyphenation{de-vices}
\RCS$Revision: 387542 $
\RCS$HeadURL: svn+ssh://svn.cern.ch/reps/tdr2/papers/SUS-14-022/trunk/SUS-14-022.tex $
\RCS$Id: SUS-14-022.tex 387542 2017-02-16 09:05:24Z paktinat $
\newcommand{\mt}{\ensuremath{{M_\mathrm{T}}}\xspace}
\newcommand{\mttwo}{\ensuremath{{M_\mathrm{ T2}}}\xspace}
\newcommand{\mindphifour}{\ensuremath{\Delta\phi_\text{min}}\xspace}
\newcommand{\Tau}{\ensuremath{\tau_\mathrm{h}}\xspace}
\newcommand{\visTau}{\ensuremath{\tau_\text{vis}}\xspace}
\newcommand{\tauMT}{\ensuremath{M_\mathrm{T}^{\Tau}}\xspace}
\newcommand{\tauTau}{\ensuremath{\Tau\Tau}\xspace}
\newcommand{\muTau}{\ensuremath{\mu\Tau}\xspace}
\newcommand{\eTau}{\ensuremath{\Pe\Tau}\xspace}
\newcommand{\leptonTau}{\ensuremath{\ell\Tau}\xspace}
\newcommand{\SumMT}{ \ensuremath{\Sigma M_\mathrm{T}^{\tau_i}}\xspace}
\newcommand{\mvisi}{\ensuremath{m^{\text{vis}(i)}}\xspace}
\newcommand{\etvisi}{\ensuremath{\et^{\text{vis}(i)}}\xspace}
\newcommand{\vptvisi}{\ensuremath{\ptvec^{\text{vis}(i)}}\xspace}
\newcommand{\wjets}{\ensuremath{\PW\text{+jets}}\xspace}
\newcommand{\genMET}{\ensuremath{\pt^\text{gen}\hspace{-1.75em}/\kern 1.0em}\xspace}
\newcommand{\binone}{SR1\xspace}
\newcommand{\bintwo}{SR2\xspace}
\newcommand{\MPT}{\ensuremath{\pt^\text{miss}}\xspace}
\newcommand{\deltaphi}{\ensuremath{\Delta\phi}\xspace}
\newcommand{\cmsTable}[1]{\resizebox{\textwidth}{!}{#1}}

\newlength\cmsFigWidth
\ifthenelse{\boolean{cms@external}}{\setlength\cmsFigWidth{0.85\columnwidth}}{\setlength\cmsFigWidth{0.4\textwidth}}
\ifthenelse{\boolean{cms@external}}{\providecommand{\cmsLeft}{top\xspace}}{\providecommand{\cmsLeft}{left\xspace}}
\ifthenelse{\boolean{cms@external}}{\providecommand{\cmsRight}{bottom\xspace}}{\providecommand{\cmsRight}{right\xspace}}
\cmsNoteHeader{SUS-14-022}
\title{Search for electroweak production of charginos in final states with two $\tau$ leptons in pp collisions at $\sqrt{s}= 8\TeV$}

\date{\today}

\abstract{Results are presented from a search for the electroweak production of supersymmetric particles in pp collisions in final states with two $\tau$ leptons. The data sample corresponds to an integrated luminosity between 18.1\fbinv and 19.6\fbinv depending on the final state of $\tau$ lepton decays, at $\sqrt{s} = 8\TeV$, collected by the CMS experiment at the LHC. The observed event yields in the signal regions are consistent with the expected standard model backgrounds. The results are interpreted using simplified models describing the pair production and decays of charginos or $\tau$ sleptons. For models describing the pair production of the lightest chargino, exclusion regions are obtained in the plane of chargino mass vs. neutralino mass under the following assumptions: the chargino decays into third-generation sleptons, which are taken to be the lightest sleptons, and the sleptons masses lie midway between those of the chargino and the neutralino. Chargino masses below  420\GeV are excluded at a 95\% confidence level in the limit of a massless neutralino, and for neutralino masses up to 100\GeV, chargino masses up to 325\GeV are excluded at 95\% confidence level. Constraints are also placed on the cross section for pair production of $\tau$ sleptons as a function of mass, assuming a massless neutralino.}

\hypersetup{%
pdfauthor={CMS Collaboration},%
pdftitle={Search for electroweak production of charginos in final states with two tau leptons in pp collisions at sqrt(s)= 8 TeV},%
pdfsubject={CMS},%
pdfkeywords={CMS, physics, supersymmetry}}

\maketitle
\section{Introduction}
\label{sect:introduction}

Supersymmetry (SUSY) \cite{Golfand:1971iw,Wess:1973kz,Wess:1974tw,Fayet1,Fayet2} is one of the most promising extensions of the
standard model (SM) of elementary particles.
Certain classes of SUSY models can lead to the unification of gauge couplings at high energy,
provide a solution to the gauge hierarchy problem without fine tuning by stabilizing the mass of the Higgs boson
against large radiative corrections, and provide a stable dark matter candidate in models with conservation of R-parity.
A key prediction of SUSY is the existence of new particles with the same gauge quantum numbers as SM particles but
differing by a half-unit in spin (sparticles).

Extensive searches at the LHC have excluded the existence of strongly produced (colored) sparticles in a broad range of scenarios,
with lower limits on sparticle masses ranging up to 1.8\TeV for gluino pair production
\cite{Chatrchyan:2013fea,Chatrchyan:2013mys,Chatrchyan:2014aea,Chatrchyan:2014lfa,Khachatryan:2015vra,Khachatryan:2015lwa,Aad:2015pfx,Aad:2015iea}.
While the limits do depend on the details of the assumed SUSY particle mass spectrum,
constraints on the colorless sparticles are generally much less stringent.
This motivates the electroweak SUSY search described in this paper.

Searches for charginos ($\PSGcpm$), neutralinos ($\PSGcz$), and sleptons ($\widetilde{\ell}$) by the ATLAS and CMS Collaborations are described in Refs.~\cite{Aad:2014nua,Aad:2014vma,Khachatryan:2014qwa,Khachatryan:2014mma,Khachatryan:2015kxa,Aad:2014yka,Aad:2015eda}.
In various SUSY models, the lightest SUSY partners of the SM leptons are those of the third generation,
resulting in enhanced branching fractions for final states with $\tau$ leptons~\cite{Martin:1997ns}.
The previous searches for charginos, neutralinos, and sleptons by the CMS Collaboration  either did not include the possibility that
the scalar $\tau$ lepton and its neutral partner (\PSGt and $\sNu_\tau$)
are the lightest sleptons \cite{Khachatryan:2014qwa}, or that the initial charginos and neutralinos are produced in vector-boson fusion processes \cite{Khachatryan:2015kxa}. An ATLAS search for SUSY in the di-$\tau$ channel is reported in Ref.~\cite{Aad:2014yka}, excluding chargino masses up to 345\GeV
for a massless neutralino (\PSGczDo).
The ATLAS results on direct \PSGt production is improved and updated in Ref.~\cite{Aad:2015eda}.

\begin{figure}[!htb]
\centering
\includegraphics[width=0.45\textwidth]{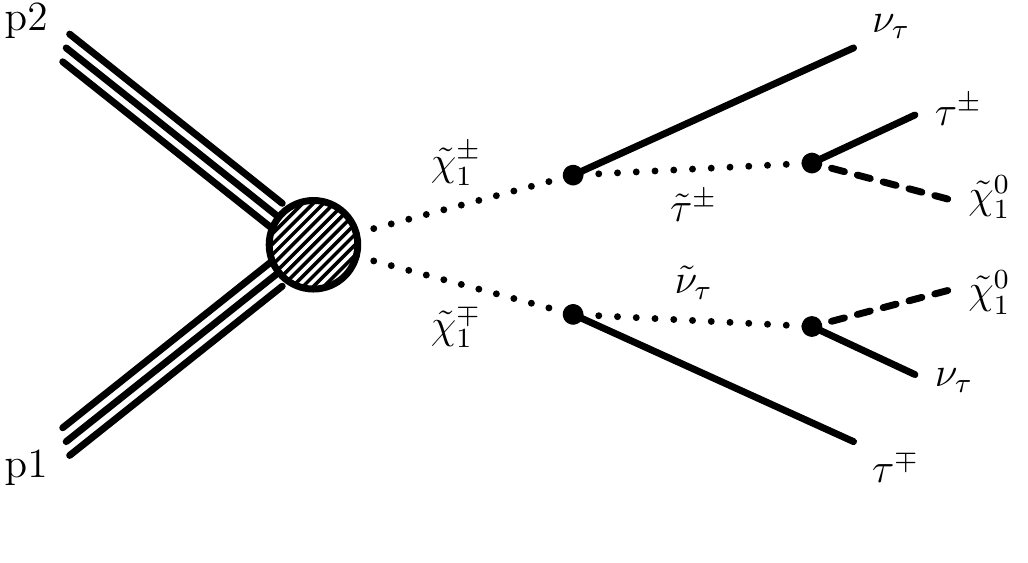}
\includegraphics[width=0.41\textwidth]{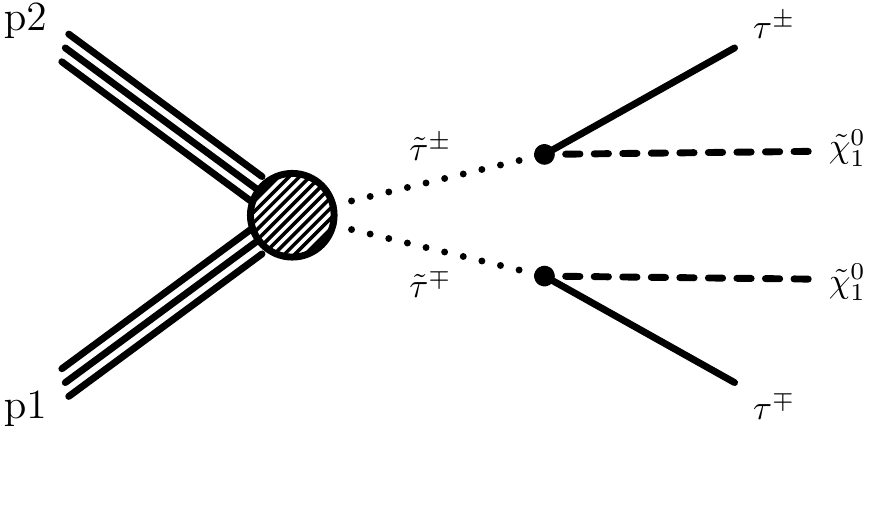}
\caption{Schematic production of $\tau$ lepton pairs from chargino (left) or $\tau$ slepton (right) pair production.}
\label{fig:Productions}
\end{figure}

In this paper, a search for the electroweak production of the lightest charginos (\PSGcpmDo) and scalar $\tau$  leptons (\PSGt) is reported using events
with two opposite-sign $\tau$ leptons and
a modest requirement on the magnitude of the missing transverse  momentum vector,  
assuming the masses of the third-generation sleptons are between those of the
chargino and the lightest neutralino.
Two $\tau$ leptons can be generated in the decay chain of \PSGcpmDo and \PSGt, as shown in Fig.~\ref{fig:Productions}.
The results of the search are interpreted in the context of SUSY simplified model spectra (SMS) \cite{Alwall:2008ag,alves:sms} for both
production mechanisms.

The results are based on a data set of proton-proton (pp)
collisions at $\sqrt{s} = 8$\TeV
collected with the CMS detector at the LHC during 2012, corresponding to integrated
luminosities of 18.1 and 19.6\fbinv in different channels.
This search makes use of the stransverse mass variable (\mttwo)~\cite{Lester:1999tx,Barr:2003rg},
which is the extension of transverse mass (\mt) to the case
where two massive particles with equal mass are created in pairs
and decay to two invisible and two visible particles.
In the case of this search, the visible particles are both $\tau$ leptons.
The distribution of \mttwo reflects the scale of the produced particles and has a longer tail for heavy sparticles
compared to lighter SM particles. Hence, SUSY
can manifest itself
as an excess of events in the high-side tail of the \mttwo distribution.
Final states are considered where
two $\tau$ leptons are each reconstructed via hadronic decays (\tauTau),
or where only one $\tau$ lepton  decays hadronically and
the other decays leptonically (\leptonTau, where $\ell$ is an electron or muon).

The paper is organized as follows.  The CMS detector, the event reconstruction, and the data sets are described
in Sections \ref{sect:CMSRec} and \ref{sect:MCSamples}. The \mttwo variable is introduced in Section \ref{sect:mt2def}.
The selection criteria for the \tauTau and \leptonTau channels are described in Section \ref{sect:tauTauCuts} and \ref{sect:eleTauCuts}, respectively.
A detailed study of the SM backgrounds is presented in Section \ref{sect:bkg}, while Section \ref{sect:sys}
is devoted to the description of the systematic uncertainties.  The results of the search with its statistical interpretation are presented in
Section \ref{sect:stat}. Section \ref{sect:conclusion} presents the summaries. The efficiencies for the important selection criteria are summarized in Appendix \ref{sect:model} and can be used to interpret these results within other phenomenological models.

\section{The CMS detector and event reconstruction}
\label{sect:CMSRec}
The central feature of the CMS apparatus is a superconducting solenoid of 6\unit{m}
internal diameter that provides a magnetic field of 3.8\unit{T}. Within the solenoid
volume are a silicon pixel and strip tracker, a lead tungstate crystal electromagnetic
calorimeter, and a brass and scintillator hadron calorimeter, each composed of a barrel
and two endcap sections. Muons are measured in gas-ionization detectors embedded in the
steel flux-return yoke outside the solenoid. Extensive forward calorimetry complements the coverage provided by the barrel and endcap detectors.
A more detailed description of the CMS detector, together with a definition of the coordinate system used
and the relevant kinematic variables, can be found in Ref. \cite{Chatrchyan:2008zzk}.

To be recorded for further study, events from pp interactions must satisfy criteria imposed by a two-level trigger system.
The first level of the CMS trigger system, composed of custom hardware processors, uses information from the
calorimeters and muon detectors to select the most interesting events in a fixed time interval of less than 4\mus.
The high-level trigger processor farm further decreases the event rate from around 100\unit{kHz} to less than 1\unit{kHz}
before data storage \cite{Khachatryan:2016bia}.

The particle-flow (PF) algorithm~\cite{CMS-PAS-PFT-09-001,CMS-PAS-PFT-10-001} reconstructs and identifies each
individual particle with an optimized combination of information from the various elements of the CMS detector.
Jets are reconstructed from the PF candidates with the anti-$k_t$ clustering
algorithm~\cite{Cacciari:2008gp} using a distance parameter of 0.5.
We apply corrections dependent on transverse momentum (\pt) and pseudorapidity ($\eta$)
to account for residual effects of nonuniform detector response~\cite{Chatrchyan:2011ds}.
A correction to account for multiple pp collisions within the same or nearby
bunch crossings (pileup interactions) is estimated on an event-by-event basis using the
jet area method described in Ref.~\cite{Cacciari:2007fd}, and is
applied to the reconstructed jet \pt.
The combined secondary vertex algorithm~\cite{Chatrchyan:2012jua} is used to identify (``b tag'') jets
originating from b quarks.  This algorithm  is based on the reconstruction of secondary vertices, together with track-based lifetime information.
In this analysis a working point is chosen such that, for jets with a \pt value greater than 60\GeV
the efficiency for tagging a jet containing a b quark is 70\% with a light-parton jet misidentification rate of 1.5\%,
and $\cPqc$ quark jet misidentification rate of 20\%.
Scale factors are applied to the simulated events to reproduce the tagging efficiencies measured in data,
separately for jets originating from b or $\cPqc$ quarks, and from light-flavor partons.
Jets with  $\pt > 40$\GeV and $\abs{\eta} < 5.0$ and b-tagged jets with $\pt > 20$\GeV and $\abs{\eta} < 2.4$ are considered in this analysis.

The PF candidates are used to reconstruct the missing transverse momentum  vector \ptvecmiss,
defined as the negative of the vector sum of the transverse momenta of all PF candidates.
For each event, \MPT is defined as the magnitude of \ptvecmiss.

Hadronically decaying $\tau$ leptons are reconstructed using the hadron-plus-strips algorithm~\cite{Khachatryan:2015dfa}.
The constituents of the reconstructed jets are used to identify individual $\tau$ lepton decay modes with one charged
hadron and up to two neutral pions, or three charged hadrons.
Additional discriminators are used to separate \Tau from electrons and muons.
Prompt $\tau$ leptons are expected to be isolated in the detector.
To discriminate them from quantum chromodynamics (QCD) jets,
an isolation variable \cite{Khachatryan:2014wca} is defined by the scalar sum of the transverse momenta of the charged hadrons 
and photons falling within 
a cone around the $\tau$ lepton momentum direction after correcting for the effect of
pileup. The ``loose'', ``medium'', and ``tight'' working points are defined
by requiring the value of the isolation variable not to exceed 2.0, 1.0,
and 0.8 \GeV, respectively.
A similar measure of isolation is computed for charged leptons (e or $\mu$),
where the isolation variable is divided by the \pt of the lepton. This quantity is
used to suppress the contribution from leptons produced in hadron decays in jets.

\section{The Monte Carlo samples}
\label{sect:MCSamples}
The SUSY signal processes and SM samples, which are used to evaluate potential background contributions,
are simulated using CTEQ6L1~\cite{Nadolsky:2008zw} parton distribution functions.
To model the parton shower and fragmentation, all generators are interfaced with \PYTHIA 6.426~\cite{Sjostrand:2006za}.
The SM processes of $\cPZ$+jets, \wjets, $\cPqt\cPaqt$, and dibosons are generated using the \MADGRAPH 5.1~\cite{Alwall:2011uj} generator.
Single top quark and Higgs boson events are generated with {\POWHEG} 1.0~\cite{Nason:2004rx,Frixione:2007vw,Alioli:2009je,Alioli:2010xd}.
In the following, the events from Higgs boson production via gluon fusion, vector-boson fusion, or in association with a $\PW$ or $\cPZ$  boson
or a \ttbar pair are referred to as ``hX.'' Later on,
the events containing at least one top quark or one $\cPZ$ boson are referred to as ``tX'' and ``ZX,'' respectively.
The masses of the top quark and Higgs boson are set to be 172.5\GeV~\cite{Khachatryan:2015hba} and 125\GeV~\cite{Aad:2015zhl}, respectively. Since the
final state arising from the pair production of \PW~bosons decaying into $\tau$ leptons is very similar
to our signal, in the following figures its contribution is shown as an independent sample labeled as ``WW.''

In one of the signal samples, pairs of charginos are produced with \PYTHIA 6.426 and decayed exclusively to the final states that contain
two $\tau$ leptons, two $\tau$ neutrinos, and two neutralinos, as shown in Fig.~\ref{fig:Productions} (left).
The daughter sparticle in the two-body decay of the \PSGcpmDo can be either a \PSGt or $\sNu_{\tau}$.
In this scenario, no decay modes are considered other than those shown in Fig.~\ref{fig:Productions} (left), so for m(\PSGt) = m($\sNu_{\tau}$), the two decay chains
(via the \PSGt or $\sNu_{\tau}$) have 50\% branching fraction.
The masses of the \PSGt and $\sNu_{\tau}$ are set to be equal to the mean value of the \PSGcpmDo and \PSGczDo masses and consequently are produced on mass shell.
If the \PSGt ($\sNu_{\tau}$) mass is close to the \PSGczDo mass, the $\tau$ lepton from the \PSGt (\PSGcpmDo) decay will
have a low (high) momentum, resulting in a lower (higher) overall event selection efficiency,
producing a weaker (stronger) limit on the chargino mass.
In the case where the \PSGt ($\sNu_{\tau}$) mass is close to the \PSGcpmDo mass, the situations are opposite.
Of the scenarios in which the $\tau$ slepton and the $\tau$ sneutrino have the same mass, the scenario with the highest efficiency overall
corresponds to the one in which these masses are half-way between the masses of the \PSGcpmDo and \PSGczDo.
In the other signal sample, pairs of staus are also produced with \PYTHIA 6.426,
that decay always to two $\tau$ leptons and two neutralinos, Fig.~\ref{fig:Productions} (right).
To improve the modeling of the $\tau$ lepton decays, the \TAUOLA 1.1.1a~\cite{Davidson:2010rw} package is used for both signal and background events.

In the data set considered in this paper,
there are on average 21 pp interactions in each bunch crossing.
Such additional interactions are generated with \PYTHIA and superimposed on simulated events in a manner consistent with
the instantaneous luminosity profile of the data set.
The detector response in the  Monte Carlo (MC) background event samples is modeled by a
detailed simulation
of the CMS detector based on {\GEANTfour}~\cite{Agostinelli:2002hh}.
For the simulation of signal events, many samples of events, corresponding to a grid of \PSGcpmDo and \PSGczDo mass values, must be generated.
To reduce  computational requirements, signal events are processed by the CMS fast simulation \cite{Abdullin:2011zz} instead of {\GEANTfour}.
It is verified that the CMS fast simulation is in reasonable agreement with the detailed simulation for our signal which has hadronic decays of
tau leptons in the final state.
The simulated events are reconstructed with similar algorithms used for collision data.

The yields for the simulated SM background samples are normalized to the cross sections available in the literature.
These cross sections correspond to next-to-next-to-leading-order (NNLO) accuracy for $\cPZ$+jets~\cite{Melnikov:2006kv}
and \wjets~\cite{xsec_WZ} events. For the $\cPqt\cPaqt$ simulated samples, the cross section used is calculated to full NNLO accuracy including
the resummation of next-to-next-to-leading-logarithmic (NNLL) terms~\cite{Czakon:2011xx}.
The event yields from diboson production are normalized to the next-to-leading-order (NLO) cross section  taken from Ref.~\cite{Campbell:2011bn}.
The \textsc{Resummino}~\cite{Fuks:2012qx,Fuks:2013vua,Fuks:2013lya} program is used to calculate the signal cross sections at NLO+NLL level where
NLL refers to next-to-leading-logarithmic precision.
\section{\texorpdfstring{Definition of \mttwo}{Definition of MT2}}
\label{sect:mt2def}
The $\mttwo$ variable~\cite{Lester:1999tx,Barr:2003rg} is used in this analysis to discriminate between the SUSY signal and the SM backgrounds as proposed in Ref \cite{Barr:2009wu}. This variable has been used extensively by both CMS and ATLAS in searches for supersymmetry \cite{Khachatryan:2015vra,Aad:2014yka}.
The variable was introduced to measure the mass of primary pair-produced  particles that eventually decay to undetected particles (e.g. neutralinos). Assuming the two primary SUSY particles undergo the same decay chain with visible and undetectable particles in the final state, the system can be described by the visible mass ($\mvisi$), transverse energy ($\etvisi$), and transverse momentum ($\vptvisi$) of each decay branch ($i=1,2$), together with the
\ptvecmiss, which is shared between the two decay chains. The quantity \ptvecmiss is interpreted as the sum of the transverse momenta
of the neutralinos, $\vec{p}_\mathrm{T}^{\PSGczDo(i)}$.
In decay chains with neutrinos, \ptvecmiss also includes contributions from the \ptvec of the neutrinos.

The transverse mass of each branch can be defined as
\begin{linenomath}
\begin{equation}
\label{eq:mtdef}
(\mt^{(i)})^{2}= (\mvisi)^2+m^2_{\PSGczDo}+2(\etvisi\et^{\PSGczDo(i)}-{\vptvisi}.\,{\vec{\pt}^{\PSGczDo(i)}}).
\end{equation}
\end{linenomath}

For a given $m_{\PSGczDo}$, the $\mttwo$ variable is defined as
\begin{linenomath}
\begin{equation}
\label{eq:mt2def}
\mttwo(m_{\PSGczDo})=\min_{\vec{p}^{\PSGczDo(1)}_\mathrm{T}+\vec{p}_\mathrm{T}^{\PSGczDo(2)}=\ptvecmiss}\,\left[\max\{ \mt^{(1)},\mt^{(2)}\}\right].
\end{equation}
\end{linenomath}

For the correct value of $m_{\PSGczDo}$, the kinematic endpoint of the $\mttwo$ distribution is at the mass of the primary particle ~\cite{Affolder:2000bpa,Abazov:2002bu}, and it shifts accordingly when the assumed $m_{\PSGczDo}$ is lower or higher than the correct value. In this analysis,
the visible part of the decay chain consists of either the two $\tau_h$ (\tauTau channel)
or a combination of a muon or an electron with a \Tau candidate ($\leptonTau$ channel), so $\mvisi$ is the mass of a lepton and can be set to zero. We also set $m_{\PSGczDo}$ to zero.

The background processes with a back-to-back topology of \tauTau  or \leptonTau 
are expected from  Drell--Yan (DY) or dijet events 
where two  jets are misidentified as \tauTau or \leptonTau. 
The resulting \mttwo value is close to zero with our choices of $m_{\PSGczDo}$ and $\mvisi$, regardless of the values of
\MPT and the \pt of  the $\tau$ candidates. 
This is not the case for signal events, where the leptons are not in a back-to-back topology because of the presence of two undetected neutralinos.
\section{\texorpdfstring{Event selection for the \tauTau channel}{Event selection for the tau-tau channel}}
\label{sect:tauTauCuts}
In this channel data of pp collisions,  corresponding to an integrated luminosity of 18.1\fbinv, are used.
The events are first selected with a trigger \cite{Chatrchyan:2011nv} that requires the presence of two isolated
\Tau candidates with $\pt > 35$\GeV and $\abs{\eta}<$ 2.1, passing loose identification requirements.
Offline, the two \Tau candidates must pass the medium $\tau$ isolation discriminator,
$\pt > 45$\GeV and $\abs{\eta}<$ 2.1, and have opposite sign (OS).
In events with more than one \tauTau pair, only the pair with the most isolated \Tau objects is considered.

Events with extra isolated electrons or muons of $\pt > 10$\GeV and $\abs{\eta} <$ 2.4
are rejected to suppress backgrounds from diboson decays.
Inspired from the MC studies,
the contribution from the $\Z \to \tauTau$  background is reduced by rejecting events
where the visible di-\Tau invariant mass is between 55 and 85\GeV (\Z boson veto).
Furthermore, contributions from low-mass DY and QCD multijet production are
reduced by requiring the invariant mass to be greater than 15\GeV.
To further reduce  $\Z\to \tauTau$ and QCD multijet events,
$\MPT > 30$\GeV and $\mttwo > 40$\GeV are also required.
The minimum angle \deltaphi in the transverse plane between the \ptvecmiss and any of the \Tau and jets,
including b-tagged jets, must be greater than 1.0 radians.
This requirement reduces backgrounds from QCD multijet events and \wjets events.

After applying the preselection described above,
additional requirements are introduced to define two search regions.
The first search region (\binone) targets models with a large mass difference ($\Delta m$)
between charginos and neutralinos.
In this case, the \mttwo signal distribution can have a long tail beyond the
distribution of SM backgrounds.
The second search region (\bintwo) is dedicated to models with small values of $\Delta m$.
In this case, the sum of the two transverse mass values, $\SumMT = \mt(\Tau^1,\ptvecmiss) + \mt(\Tau^2,\ptvecmiss)$,
provides additional discrimination between signal and SM background processes.

The two signal regions (SR) are defined as:
\begin{itemize}
\item {\binone}: $\mttwo > 90$\GeV;
\item {\bintwo}:  $\mttwo < 90$\GeV, $\SumMT > 250$\GeV, and events with b-tagged jets are vetoed.
\end{itemize}
The veto on events containing b-tagged jets in \bintwo reduces the number of \ttbar events, which are expected in  the low-\mttwo region.
Table \ref{Tab.Cuts} summarizes the selection requirements for the different signal regions.
\section{\texorpdfstring{Event selection for the \leptonTau channel}{Event selection for the lepton-tau channel}}
\label{sect:eleTauCuts}
Events in the \leptonTau final states (e\Tau and $\mu\Tau$)
are collected with triggers that require
a loosely isolated \Tau with $\pt > 20$\GeV and $\abs{\eta}< 2.3$, as well as
an isolated electron or muon with $\abs{\eta} < 2.1$ \cite{Chatrchyan:2011nv,Khachatryan:2015hwa,Chatrchyan:2012xi}.  The minimum
\pt requirement for the electron (muon) was increased during the data taking from 20 to 22\GeV (17 to 18\GeV)
due to the increase in instantaneous luminosity. An integrated luminosity of 19.6\fbinv is used to study these channels.

In the offline analysis, the electron, muon, and \Tau objects are required to have $\pt > 25$, 20, and 25\GeV, respectively,
and the corresponding identification and isolation requirements are tightened. The $\abs{\eta}$ requirements are the same as those in the online selections.
In events with more than one opposite-sign \leptonTau pair, only
 the pair that maximizes the scalar $\pt$ sum of \Tau and electron or muon is considered.  Events with additional loosely isolated leptons
with $\pt > 10$\GeV are rejected to suppress backgrounds from $\Z$ boson
decays.

Just as for the \tauTau channel, preselection requirements to suppress
QCD multijet, \ttbar, $\Z \to \tau \tau$, and low-mass resonance events are applied.
 These requirements are \leptonTau 
invariant mass between 15 and 45\GeV or $>$ 75\GeV (\Z boson veto), $\MPT > 30$\GeV, $\mttwo > 40$\GeV, 
and $\deltaphi > 1.0$ radians. 
The events with b-tagged jets are also rejected to reduce the \ttbar background.
  The final signal region requirements are $\mttwo > 90$\GeV and $\tauMT > 200$\GeV.
The latter requirement provides discrimination against the \wjets background.  Unlike in the \tauTau channel,
events with $\mttwo < 90$\GeV are not used because of the higher
level of background.

\begin{figure}[!htb]
\centering
\includegraphics[width=0.475\textwidth]{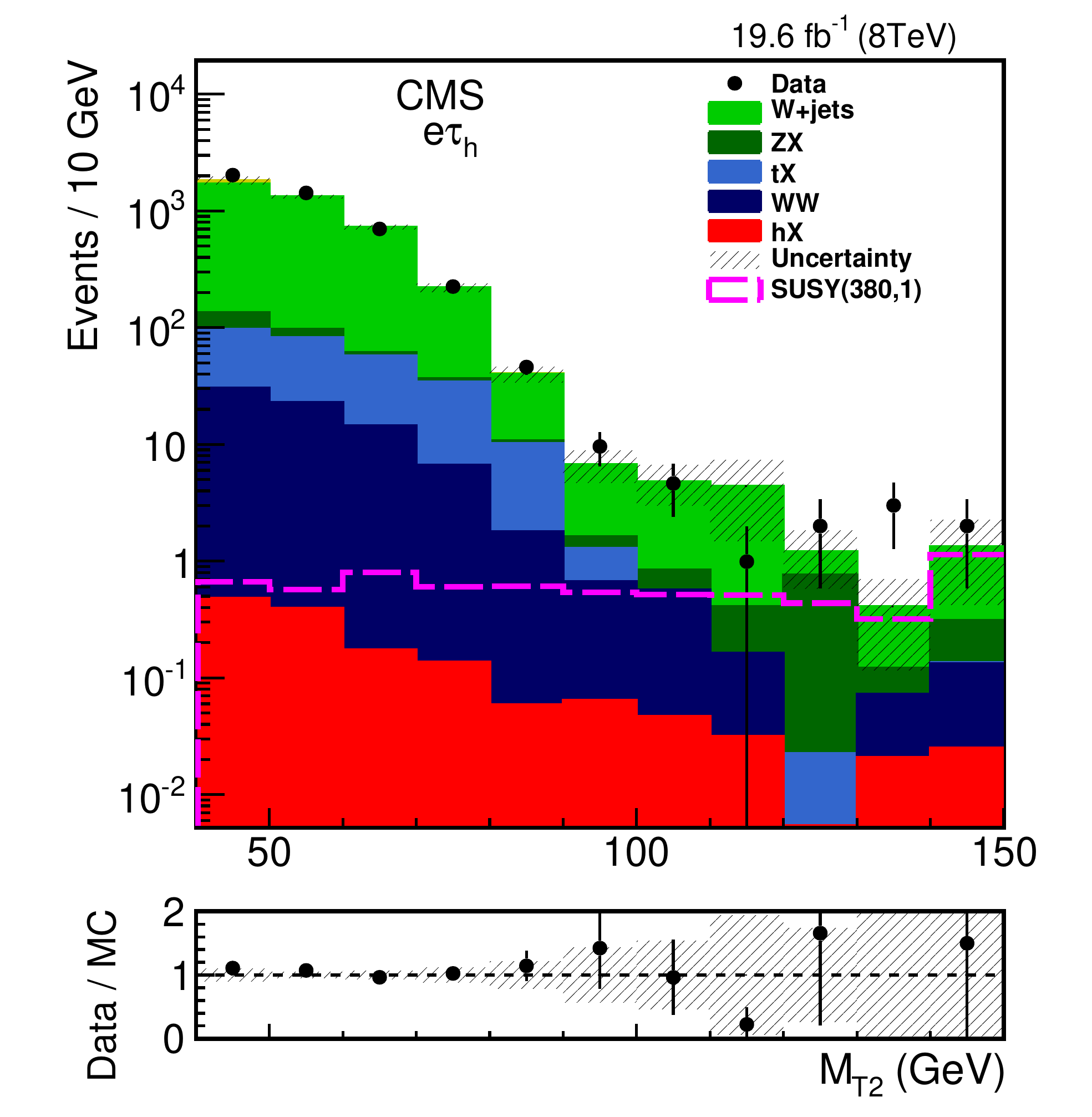}
\includegraphics[width=0.475\textwidth]{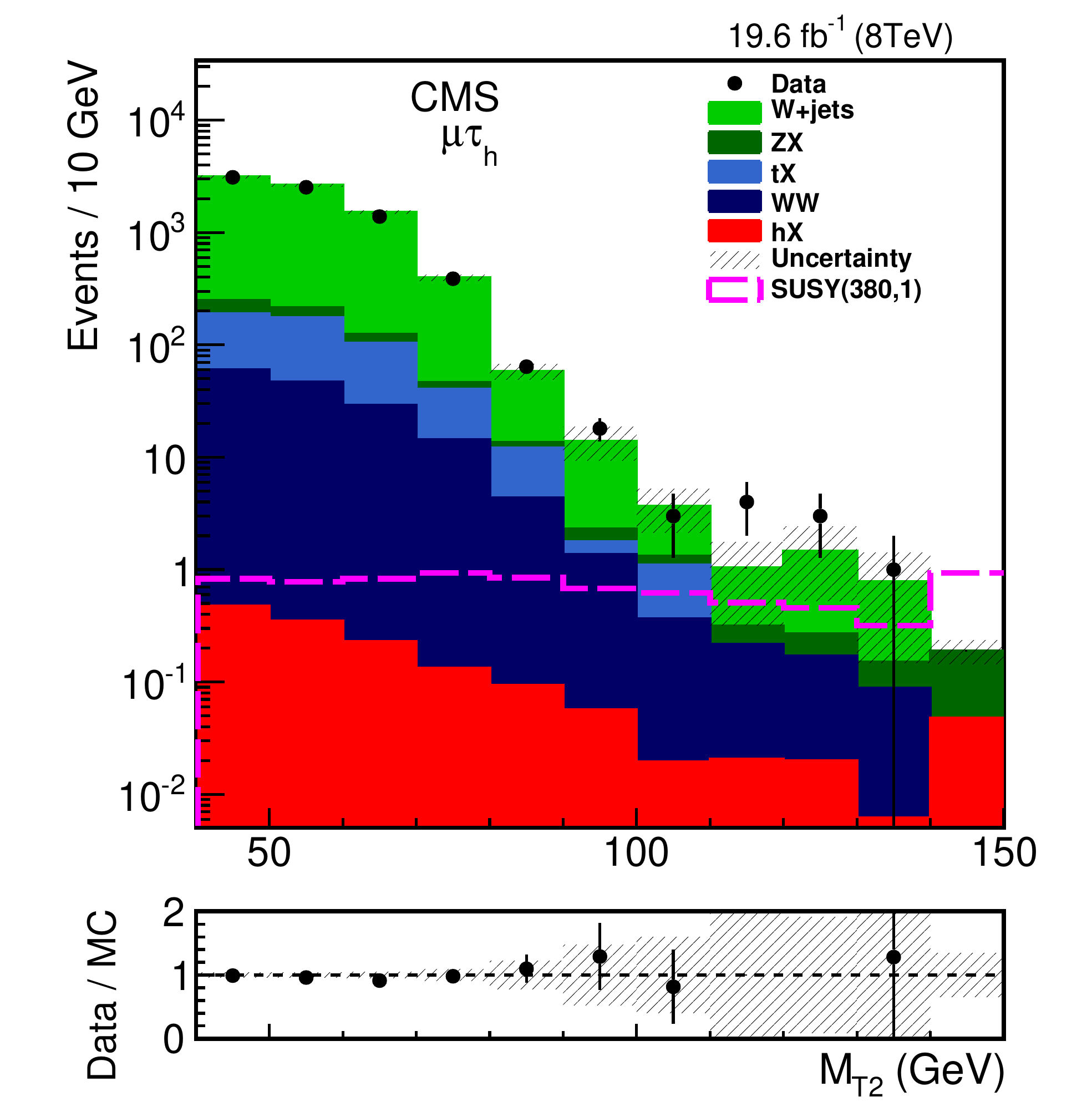}
\caption{The \mttwo distribution before applying the final selections on \mttwo and \tauMT, compared to SM expectation in (left) \eTau and (right) \muTau channels. The signal distribution is shown for $m_{\PSGcpmDo}=380\GeV, m_{\PSGczDo}=1\GeV$. The last bins include all overflows to higher values of \mttwo. Only the statistical uncertainties are shown.}
\label{fig:mt2leptontau}
\end{figure}

The summary of the selection requirements is shown in Table \ref{Tab.Cuts}.
\begin{table}[!htb]
\centering
\topcaption{Definition of the signal regions.}
\begin{tabular}{c|c|c}
\hline
   \leptonTau& \tauTau & \tauTau               \\
             & \binone & \bintwo               \\\hline\hline
 OS \leptonTau & \multicolumn{2}{c}{OS \tauTau}  \\\hline
\multicolumn{3}{c}{Extra lepton veto}          \\
\multicolumn{3}{c}{Invariant mass of \leptonTau or $\tauTau > 15$\GeV}\\
\multicolumn{3}{c}{\Z boson mass veto}              \\
\multicolumn{3}{c}{$\MPT > 30$\GeV}            \\
\multicolumn{3}{c}{$\mttwo > 40\GeV$}         \\
\multicolumn{3}{c}{$\deltaphi > 1.0\unit{radians}$}         \\\hline
\multicolumn{1}{l|}{b-tagged jet veto}&  --- & \multicolumn{1}{l}{b-tagged jet veto}  \\
\multicolumn{2}{c|}{~~~~~$\mttwo > 90\GeV$} & \multicolumn{1}{l}{$\mttwo < 90\GeV$} \\
\multicolumn{1}{l|}{$\tauMT > 200\GeV$}    &  --- & \multicolumn{1}{l}{$\SumMT > 250\GeV$} \\\hline
\end{tabular}
\label{Tab.Cuts}
\end{table}
Figure \ref{fig:mt2leptontau}
shows the \mttwo distribution after the preselection requirements are imposed.
The data are in good agreement with the SM expectations, evaluated from MC simulation, within the statistical uncertainties.
A SUSY signal corresponding to high $\Delta m$ ($m_{\PSGcpmDo}=380\GeV,~m_{\PSGczDo}=1\GeV)$ is used to show the expected signal distribution.

\section{Backgrounds}
\label{sect:bkg}
The backgrounds are studied in two categories: those with
``misidentified'' \Tau, \ie, events where a quark or gluon jet has been misidentified
as a \Tau, and those with genuine \Tau candidates.
The QCD multijet and \wjets events are the dominant sources in the first category, while a mixture of \ttbar, Z+jets, diboson, and Higgs boson
events dominate the second category. Background estimates are performed using control samples in data whenever possible.
Those backgrounds that are taken from simulation are either validated in dedicated control regions or corrected using data-to-simulation scale factors.
The estimates of the main backgrounds are discussed below, while the remaining contributions are small and are taken from simulation.

\subsection{\texorpdfstring{The QCD multijet background estimation in the \tauTau channel}{The QCD multijet background estimation in the tau-tau channel}}
\label{sect:bkgQCD}
\begin{figure}[!htb]
\centering
\includegraphics[angle=0,scale=1.15]{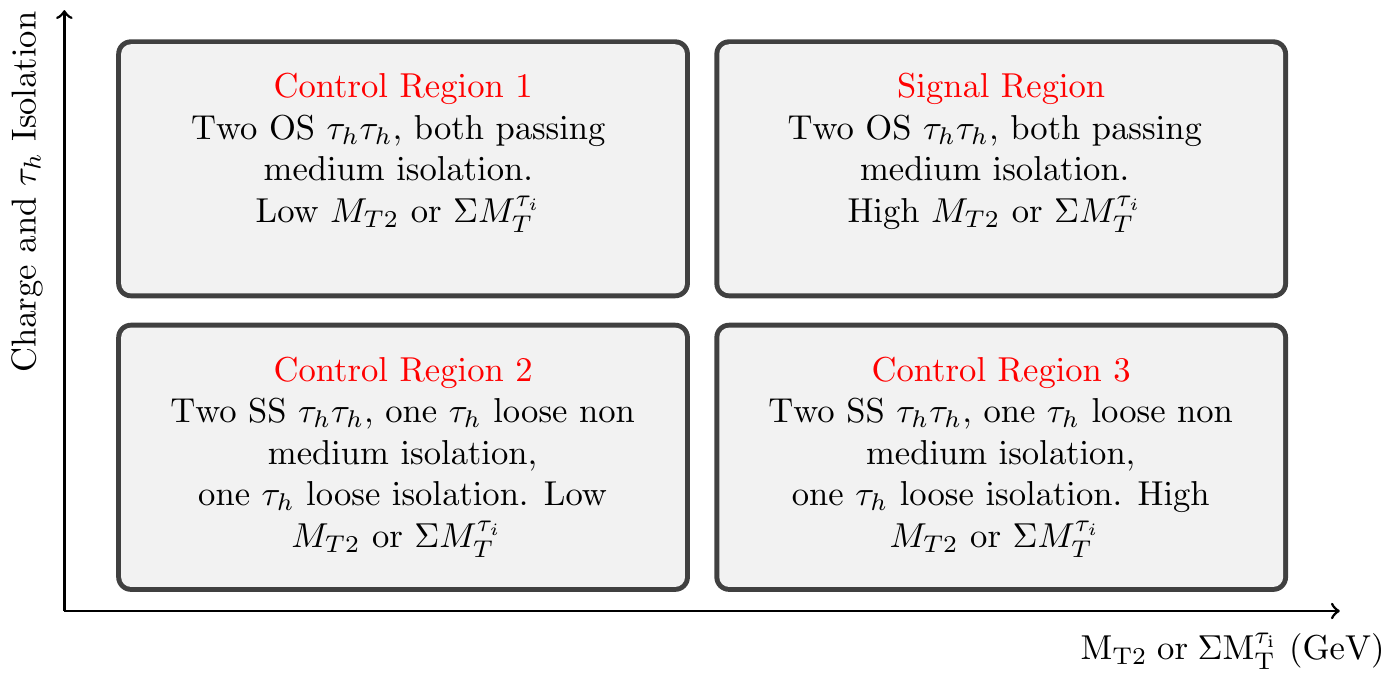}
\caption{Schematic illustration of three control regions and the signal region used to estimate the QCD multijet background.}
\label{fig:ABCDQCD}
\end{figure}

Events from QCD multijet production can appear in the signal regions if two hadronic jets are misidentified as a \tauTau pair.
The isolation variable is a powerful discriminant between misidentified and genuine \Tau candidates. To estimate the QCD multijet contribution, an ABCD method is used, where three \tauTau control regions (CRs) are defined using the loose \Tau isolation requirement, together with lower thresholds on \mttwo or \SumMT variables for the corresponding signal region. The former is changed from $\mttwo > 90$ to $>$40\GeV, whereas the latter is reduced from $\SumMT > 250$ to $>$100\GeV. In addition, the requirement on \deltaphi is removed to increase the number of events in the CRs.
To reduce contamination from genuine \tauTau events
in CRs with at least one loose \Tau candidate,
same-sign (SS) \tauTau pairs are selected. Residual contributions from genuine
\tauTau and \wjets events (non-QCD events) are subtracted based on MC expectations.
The CR and signal region are illustrated in Fig.~\ref{fig:ABCDQCD}.
In the samples dominated by QCD multijet events (CR1 and CR2), the isolation of misidentified \Tau candidates is found 
to be  uncorrelated with the search variables \mttwo and \SumMT.
The QCD multijet background in the signal regions is therefore estimated by scaling the number of QCD multijet events with high \mttwo or high \SumMT and loosely isolated SS \tauTau (CR3) by a transfer factor,
which is the $y$-intercept of a horizontal line fitted to the ratio of the numbers of events in CR1 and CR2 in different bins of the
low values of the search variables.
The final estimate of the background is corrected for the efficiency of
the \deltaphi requirement for QCD multijet events. This efficiency is measured in CR1 and CR2,
in which the contribution of QCD multijet events is more than 80\%. It is checked that the efficiency versus the search variable is
same in both CR1 and CR2 and to gain in statistics, two CRs are combined before measuring the efficiency.
The efficiency is a falling distribution as a function of
the search variable (\mttwo or \SumMT) and the value of the last bin ($65<\mttwo<90\GeV$ or $200<\SumMT<250\GeV$)
is used conservatively as the value of the efficiency in the signal regions.

The number of data events in CR3 after subtracting the non-QCD events is 4.81 $\pm$ 2.57 (8.62 $\pm$ 3.55) for the \binone (\bintwo) selection.
For SR1 (SR2), the transfer factors and  \deltaphi efficiencies are measured to be 0.91 $\pm$ 0.12 (0.89 $\pm$ 0.11) and 0.03 $^{+0.04} _{-0.03}$ (0.15 $\pm$ 0.08),
respectively.
The reported uncertainties are the quadratic sum of the statistical and systematic uncertainties.

The systematic uncertainty in the background estimates includes the uncertainty in the validity of the assumption that isolation
and \mttwo or \SumMT are not correlated, the \deltaphi efficiency is extrapolated correctly to the signal regions, and the uncertainties in the residual
non-QCD SM backgrounds which  are subtracted based on MC expectations for different components of the background estimation.
The latter includes both the statistical uncertainty of the simulated
events and also a 22\% systematic uncertainty that will be discussed in Section~\ref{sect:sys},
assigned uniformly to all simulated events.

Table \ref{4QCDbg} summarizes the estimation of the QCD multijet background contribution in the two signal regions after extrapolation from
the control regions and correcting for the \deltaphi efficiency.
To evaluate the uncertainties in the transfer factor and \deltaphi
efficiency due to the correlation
assumptions, different fit models are examined: (i) a horizontal line or a
line with a constant slope is fitted in the distributions of the transfer
factor or \deltaphi efficiency for $40<\mttwo<90\GeV$ in the \binone case ($100<\SumMT<250\GeV$ in the \bintwo case); 
or (ii) the value of the last bin adjacent to the signal region is used.
The weighted average of the estimates is compared with the reported values 
in Table \ref{4QCDbg} to extract the ``fit'' uncertainty.
\begin{table}[!htb]
\centering
\topcaption{The estimated QCD multijet background event yields in the \tauTau channel. The first two uncertainties are the statistical and systematic uncertainties of the method, and the last uncertainty is the extra systematic uncertainty due to the correlation assumptions.}
\begin{tabular}{ll}
\hline
 Signal region       & QCD multijet  background estimate\\
\hline
\tauTau \binone      & $0.13 \pm0.06\stat {}^{+0.18} _{-0.13}\syst \pm 0.10\,\text{(fit)}$ \\
\tauTau \bintwo      & $1.15 \pm 0.39\stat \pm 0.70\syst \pm 0.25\,\text{(fit)}$ \\
\hline
\end{tabular}
\label{4QCDbg}
\end{table}

\subsection{\texorpdfstring{\wjets background estimation in the \tauTau channel}{W+jets background estimation in the tau-tau channel}}
\label{sect:bkgW}
{\tolerance=1200
In the \tauTau channel, the number of remaining events for \wjets from MC is zero,
but it has a large statistical uncertainty due to the lack of the statistics in the simulated sample. To
have a better estimation, the contribution of the \wjets background in the \tauTau channel is taken from simulated events, using the formula:
\begin{equation}
N_\mathrm{SR} = \epsilon_{\mathrm{FS}}N_\mathrm{BFS} .
\end{equation}
Here $N_\mathrm{SR}$ is the estimation of \wjets events in the signal region, $N_\mathrm{BFS}$ is the number of
\wjets events before applying the final selection criterion ($\mttwo > 90$\GeV for \binone and $\SumMT > 250$\GeV for \bintwo), but after applying all other selection criteria, including $\mttwo > 40$\GeV for \binone and $40 <\mttwo< 90$\GeV for \bintwo.
The efficiency of the final selection ($\epsilon_{\mathrm{FS}}$) is defined as ${N (M_{\mathrm{T}2}>90)}/{N ( M_{\mathrm{T}2}>40)}$ for \binone and ${N (\SumMT>250)}/{N(40<M_{\mathrm{T}2}<90)}$ for \bintwo.
The value of $N_\mathrm{BFS}$ is $31.9\pm6.4\,(29.1\pm6.2)$ for \binone (\bintwo), where the uncertainties arise from the limited number of simulated events.
\par}

The $\epsilon_\mathrm{FS}$ is evaluated in a simulated \wjets sample with a pair of opposite-sign \Tau candidates, where the \Tau candidates
are selected with the same identification requirements as in the signal region, but with looser kinematic selection criteria to improve statistical precision.
Additional signal selection requirements on \deltaphi or the lepton veto are applied one by one such that two orthogonal subsamples (passing and failing) are obtained. The $\epsilon_{\mathrm{FS}}$ quantity is calculated in all subsamples.
The values are consistent with those obtained from the sample defined with relaxed requirements
within the statistical uncertainties.
The measured $\epsilon_{\mathrm{FS}}$ values from the looser-selection samples are
0.028 $\pm$ 0.010 and 0.098 $\pm$ 0.032 for \binone and \bintwo, respectively.
The uncertainty in the \Tau energy scale is also taken  into account in the uncertainty in $\epsilon_{\mathrm{FS}}$.

The \wjets simulated sample is validated in data using a same-sign \muTau control sample, where both the normalization and $\epsilon_{\mathrm{FS}}$ are checked.
The ratio of data to MC expectation is found to be $1.05 \pm0.13\,(1.02 \pm 0.09)$ for \binone(\bintwo),
which is compatible with unity within the uncertainties.
For $\epsilon_{\mathrm{FS}}$, to take into account the difference between the data and MC values,
the MC prediction in each of the two signal regions is corrected by the ratio of $\epsilon_\mathrm{FS(data)}$ to $\epsilon_\mathrm{FS(MC)}$,
which is $0.73 \pm 0.57\,(1.49 \pm 0.38)$ for \binone(\bintwo), and its uncertainty is also taken to be the ``shape'' systematic uncertainty.

Table \ref{tbl:Wbkg} summarizes the estimated results for different signal regions for the \tauTau channel.
\begin{table}[!htb]
\centering
\topcaption{The \wjets background estimate in the two search regions.
The systematic uncertainty ``syst'' comes from the maximum
variation of the estimation found  from varying the \Tau energy scale within its uncertainty.
The ``shape'' uncertainty takes into account the difference between the shape of the search variable distribution in data and simulation.}
\begin{tabular}{ll}
\hline
Signal Region & \wjets background estimate\\
\hline
\tauTau \binone & 0.70 $\pm$ 0.21\stat $\pm$ 0.09\syst $\pm$ 0.54 (shape)\\
\tauTau \bintwo & 4.36 $\pm$ 1.05\stat $\pm$ 1.14\syst $\pm$ 1.16 (shape)\\
\hline
\end{tabular}
\label{tbl:Wbkg}
\end{table}

\subsection{The Drell--Yan background estimation}
\label{sect:bkgDY}
The DY background yield is obtained from the MC simulation.
The simulated sample includes production of different lepton pairs (\Pe\Pe, $\mu\mu$, and $\tau\tau$).
The contribution from $\Z\to \ell \ell$ and $\Z\to \tau \tau\to \ell \ell$ events is found to be very small, because the misidentification probabilities for $\ell\to\Tau$ are sufficiently low.
The dominant background events are $\Z\to \tau \tau\to \ell \Tau$ and $\Z\to \tau \tau\to \Tau \Tau$ decays.
The misidentification probability for  $\Tau \to\ell$ is also low, so the probability
to have DY background contribution from $\Z\to \tau \tau\to \Tau \Tau$ events in the \leptonTau channels is negligible.
The simulation is validated in a \muTau control region obtained by removing the \deltaphi
requirement and by inverting the \Z boson veto and also by requiring $\mttwo < 20$\GeV,  $40 < \tauMT < 100$\GeV.
The distributions of the invariant mass of the \muTau system for data and simulated events are in good agreement.
The \pt of the \Z boson system, which is correlated with
\mttwo, is also well reproduced in simulation. Table \ref{tbl:DYbkg}
summarizes the DY background contribution in the different signal regions.
For \leptonTau channels, only the contributions from the genuine lepton+\Tau are reported.
A separate method is developed in Section~\ref{sect:bkgFake} to estimate the misidentified lepton contamination in these channels. The systematic uncertainties
of the DY background are discussed in detail in Section \ref{sect:sys}.
\begin{table}[!htb]
\centering
\newcolumntype{x}{D{,}{\,\pm\,}{4.8}}
\topcaption{The DY background contribution estimated from simulation in four signal regions.  The uncertainties are due to the limited number of MC events.}
\begin{tabular}{lx}
\hline
Signal Region      & \multicolumn{1}{c}{ DY background estimate}\\
\hline
\eTau              & 0.19,0.04\\
\muTau             & 0.25,0.06\\
\tauTau \binone    & 0.56,0.07\\
\tauTau \bintwo    & 0.81,0.56\\
\hline
\end{tabular}
\label{tbl:DYbkg}
\end{table}

\subsection{\texorpdfstring{Misidentified \Tau in the \leptonTau channels}{Misidentified tau in the lepton-tau channels}}
\label{sect:bkgFake}
The contribution from misidentified \Tau in the \leptonTau channels is estimated using a method which takes into account the probability
that a loosely isolated misidentified or genuine \Tau passes the tight isolation requirements.
If the signal selection is done using the \Tau candidates that pass the loose isolation,
the number of loose \Tau candidates ($N_\mathrm{l}$) is:
\begin{equation}
N_\mathrm{l} = N_\mathrm{g} + N_\mathrm{m}
\end{equation}
where $N_\mathrm{g}$ is the number of genuine \Tau candidates and $N_\mathrm{m}$ is the number of misidentified
\Tau candidates. If the selection is tightened, the number of tight \Tau candidates ($N_\mathrm{t}$)  is
\begin{equation}
 N_\mathrm{t} = r_\mathrm{g} N_\mathrm{g} + r_\mathrm{m} N_\mathrm{m}
\end{equation}
where $r_\mathrm{g}$ ($r_\mathrm{m}$) is the genuine (misidentified \Tau) rate, \ie, the probability that a loosely selected genuine (misidentified) \Tau candidate passes the  tight  selection.
One can obtain the following expression by eliminating $N_\mathrm{g}$:
\begin{equation}
   r_\mathrm{m} N_\mathrm{m}  = r_\mathrm{m} (N_\mathrm{t} - r_\mathrm{g} N_\mathrm{l})/(r_\mathrm{m}-r_\mathrm{g}).
\end{equation}
Here, the product $r_\mathrm{m} N_\mathrm{m}$ is the contamination of misidentified \Tau candidates in the signal region.
This is determined by measuring $r_\mathrm{m}$ and $r_\mathrm{g}$ along with the number of loose \Tau candidates ($N_\mathrm{l}$) and the number of tight \Tau candidates ($N_\mathrm{t}$).

The misidentification rate ($r_\mathrm{m}$) is measured as the ratio of tightly selected \Tau candidates to loosely
selected \Tau candidates in a sample dominated by misidentified \Tau candidates.
This is done in a data sample with the same selection as \leptonTau, except with an inverted
\MPT requirement, \ie, $\MPT < 30$\GeV. The misidentification rate is measured to be $0.54 \pm0.01$.
The genuine \Tau candidate rate ($r_\mathrm{g}$) is estimated in simulated DY events; it is found to
be $r_\mathrm{g} = 0.766 \pm 0.003$ and almost independent of \mttwo.
A relative systematic uncertainty of 5\% is assigned to the central value of $r_\mathrm{g}$ to cover its
variations for different values of \mttwo.
The method is validated in the simulated \wjets sample using the misidentification rate which is evaluated with the same method as used for data.
This misidentification rate is $r_\mathrm{m} = 0.51$.
This difference is taken as the systematic uncertainty of 5\%
in the central value of the misidentification rate ($r_\mathrm{m} = 0.54$).
The method predicts the number of \leptonTau background events in this sample within the
uncertainties.
These include statistical uncertainties due to the number of events in the
sidebands (loosely selected \Tau candidates), as well as
systematic uncertainties.
The uncertainties in the
misidentification rate and the genuine \Tau candidate rate
are negligible compared to the statistical uncertainties associated to
the control regions.

The estimates of the misidentified \Tau contamination in the two \leptonTau
channels are summarized in Table~\ref{Tab.FakeEstimation}.
The relative statistical and systematic uncertainties are reported separately.
Since the same misidentified and genuine \Tau candidate rates are used to estimate the backgrounds for both the
\eTau and \muTau channels, the total systematic uncertainties are considered
fully correlated between the two channels.
The numbers of misidentified events (3.30 for the \eTau channel and 8.15 for the \muTau channel) are consistent within the statistical uncertainties in our control samples. 
\begin{table}[!htb]
\centering
\topcaption{Estimation of the misidentified \Tau contribution in the signal region of the \leptonTau channels. The total systematic uncertainty is the
quadratic sum of the individual components. All uncertainties are relative.
The $r_\mathrm{m}$ ($r_\mathrm{g}$) is shorthand for misidentified (genuine) \Tau candidate rate.}
\begin{tabular}{lccccc}
\hline
Channel    & Total misid (events) & Stat (\%) &  $r_\mathrm{m}$ syst (\%) & $r_\mathrm{g}$  syst (\%) & Total uncert (\%) \\\hline
\eTau      &   3.30     &  101    &  17    & 2  & 102  \\
\muTau     &   8.15     &   56    &  18    & 5   & 59  \\
\hline
\end{tabular}
\label{Tab.FakeEstimation}
\end{table}
\section{Systematic uncertainties}
\label{sect:sys}
Systematic uncertainties can affect the shape or normalization of the
backgrounds estimated from simulation (\ttbar, Z+jets, diboson, and Higgs boson events),
as well as the signal acceptance.
Systematic uncertainties of other background contributions are described in Sections \ref{sect:bkgQCD}, \ref{sect:bkgW} and \ref{sect:bkgFake}.
The uncertainties are listed below, and summarized in Table~\ref{Tab.SYS}.

\begin{itemize}
\item  The energy scales for electron, muon, and \Tau objects affect the shape of the kinematic distributions.
 The systematic uncertainties in the muon and electron energy scales are negligible.
The visible energy of \Tau object in the MC simulation is scaled up and down
by 3\%, and all \Tau-related variables are recalculated. The resulting variations in
final yields are taken as the systematic uncertainties. They are evaluated to be 10--15\% for
backgrounds and 2--15\% in different parts of the signal phase space.

\item The uncertainty in the \Tau identification efficiency is 6\%. The uncertainty in the trigger
efficiency of the \Tau part of the \eTau and \muTau (\tauTau) triggers amounts to 3.0\% (4.5\%) per
\Tau candidate. A ``tag-and-probe'' technique \cite{Chatrchyan:2014mua} on $\cPZ\to \Pgt\Pgt$ data
events is used to estimate these uncertainties \cite{Khachatryan:2014wca}.

\item The uncertainty in electron and muon trigger, identification, and
isolation efficiencies is 2\% \cite{Khachatryan:2014wca}.

\item The uncertainty due to the scale factor for the b-tagging efficiency and misidentification
rate is evaluated by varying the factors within their uncertainties. The yields of signal and background events are changed by 8\%
and 4\%, respectively \cite{Chatrchyan:2012jua}.

\item To evaluate the uncertainty due to pileup, the measured inelastic pp cross section is
  varied by 5\% \cite{Antchev:2011vs}, resulting in a change in the number of simulated pileup interactions.
 The relevant efficiencies for signal and background events are changed by 4\%.

\item The uncertainty in the signal acceptance due to parton distribution function (PDF) uncertainties
  is taken to be 2\% from a similar analysis \cite{Khachatryan:2014qwa} which follows the PDF4LHC recommendations \cite{pdf4lhc}.

\item The uncertainty in the integrated luminosity  is 2.6\% \cite{CMS-PAS-LUM-13-001}.  This affects only the
  normalization of the signal MC samples. Because for the backgrounds  either control samples in data are used or the normalization is measured from data.

\item The uncertainty in the signal acceptance associated with initial-state radiation (ISR)
is evaluated by comparing the efficiencies of jet-related requirements
in the \MADGRAPH{}+\PYTHIA program.
Using the SM WW process, which
 is expected to be similar to chargino pair production in terms of parton content and process, a 3\% uncertainty in
the efficiency of  b-tagged jets veto and a 6\% uncertainty in the \deltaphi requirement are assigned.

\item The uncertainties related to \MPT can arise from different sources, e.g.  the energy scales of lepton, \Tau, and jet
objects, and unclustered energy.  The unclustered energy is the energy of the reconstructed objects which
 do not belong to any jet or lepton with $\pt > 10$\GeV. The effect of lepton and \Tau
 energy scales is discussed above. The contribution from the uncertainty in the jet energy scale (2--10\% depending on $\eta$  and \pt) and
 unclustered energy (10\%) is found to be negligible. A conservative value of 5\% uncertainty
 is assigned to both signal and background processes based on MC simulation studies \cite{Khachatryan:2015kxa, Khachatryan:2014qwa}.

\item The performance of the fast detector simulation has some differences compared to the full detector simulation, especially in
 track reconstruction \cite{Khachatryan:2015kxa} that can affect the \Tau isolation. A 5\% systematic uncertainty per
 \Tau candidate is assigned by comparing the \Tau isolation and identification efficiency in the fast
 and full simulations.

\item The statistical uncertainties due to limited numbers of simulated events also contributes to the overall uncertainties.
This uncertainty amounts to 3--15\% for the different parts of the signal phase space and 13--70\% for the backgrounds in different signal regions.

\item For less important backgrounds like \ttbar,  dibosons, and Higgs boson production, the number of simulated
events remaining after event selection is very small. A 50\% uncertainty is considered for these backgrounds
to account for the possible theoretical uncertainty in the cross section calculation as well as the shape mismodeling.
\end{itemize}

\begin{table}[!htb]
\centering
\topcaption{Summary of the systematic uncertainties that affect the signal event
selection efficiency, DY and  rare backgrounds normalization and their shapes.
The sources that affect the shape are indicated by (*) next to their names.
These sources are considered correlated between two signal regions of the
\tauTau analysis in the final statistical combination.}
{
\begin{tabular}{l|ccc|ccc}
\hline
Systematic uncertainty source &\multicolumn{3}{c|}{Background (\%)}         &\multicolumn{3}{c}{Signal (\%)}\\\hline
                              &   \leptonTau         & \tauTau & \tauTau         &   \leptonTau         & \tauTau & \tauTau\\
 &  & \binone &  \bintwo        &  & \binone &  \bintwo        \\
\hline
\Tau energy scale (*)          &10 &\multicolumn{2}{c|}{15}  & 2--12 &\multicolumn{2}{c}{3--15} \\
\Tau identification efficiency & 6 &\multicolumn{2}{c|}{12} & 6 &\multicolumn{2}{c}{12}  \\
\Tau trigger  efficiency       & 3&\multicolumn{2}{c|}{9}& 3&\multicolumn{2}{c}{9}  \\
Lepton trigger and ident. eff. & 2 & \multicolumn{2}{c|}{---} & 2 &  \multicolumn{2}{c}{---} \\
b-tagged jets veto              & 4 & --- & 4 &  8 & --- & 8 \\
Pileup&\multicolumn{3}{c|}{4} &\multicolumn{3}{c}{4} \\
PDF (*)&\multicolumn{3}{c|}{---}&\multicolumn{3}{c}{2} \\
Integrated luminosity       &\multicolumn{3}{c|}{---} & \multicolumn{3}{c}{2.6}\\
ISR (*)&\multicolumn{3}{c|}{---}&\multicolumn{3}{c}{3} \\
\mindphifour&\multicolumn{3}{c|}{---}&\multicolumn{3}{c}{6} \\
\MPT (*)&\multicolumn{3}{c|}{5} &\multicolumn{3}{c}{5} \\
Fast/full \Tau ident. eff. &\multicolumn{3}{c|}{---}& 5 & \multicolumn{2}{c}{10}\\\hline
Total shape-affecting sys. & 11 & 16 & 16 & 6--13 &\multicolumn{2}{c}{7--16} \\
Total non-shape-affecting sys. & 9 & 16 & 16 & 14 &20& 21 \\
Total systematic &  14 & 22  & 22& 15--19 & 21--25  & 22--26\\
MC statistics & 22 & 13 & 70 & \multicolumn{3}{c}{3--15} \\\hline
Total& 26 & 26  & 73& 15--24 & 21--29  & 22--30\\\hline
Low-rate backgrounds &\multicolumn{3}{c|}{50}&\multicolumn{3}{c}{---}\\\hline \end{tabular}
}
\label{Tab.SYS}
\end{table}

The systematic uncertainties that can alter the shapes are added in quadrature and
treated as correlated when two signal regions of the \tauTau channel are combined. Other systematic uncertainties of these two
channels and all of the systematic uncertainties of the \leptonTau channels are treated as uncorrelated.

\section{Results and interpretation}
\label{sect:stat}
The observed data and predicted background yields for the four signal regions are summarized in Table~\ref{tbl:yieldSysSummary}.
\begin{table}[!htb]
\centering
\topcaption{Data yields and background predictions with uncertainties in the four signal regions of the search.
The uncertainties are reported in two parts, the statistical and systematic uncertainties, respectively.
The \wjets and QCD multijet main backgrounds are derived from data as described in Section~\ref{sect:bkg};
the abbreviation ``VV'' refers to diboson events. The yields for three signal points representing the low, medium, and high $\Delta m$
are also shown. SUSY(X, Y) stands for a SUSY signal with ${m}_{\PSGcpmDo}= \mathrm{X}$\GeV and ${m}_{\PSGczDo} = \mathrm{Y}$\GeV.}
\cmsTable{
\begin{tabular}{lcccc}
\hline
	           & \eTau & \muTau & \tauTau \binone & \tauTau \bintwo \\
\hline
  DY               & 0.19 $\pm$ 0.04 $\pm$ 0.03 & 0.25 $\pm$ 0.06  $\pm$ 0.04  &  0.56 $\pm$ 0.07 $\pm$ 0.12 & 0.81 $\pm$ 0.56 $\pm$ 0.18  \\
tX, VV, hX  & 0.03 $\pm$ 0.03 $\pm$ 0.02 & 0.19 $\pm$ 0.09  $\pm$ 0.09  &  0.19 $\pm$ 0.03 $\pm$ 0.09 & 0.75 $\pm$ 0.35 $\pm$ 0.38  \\
\wjets             & 3.30$_{- 3.30}^{+ 3.35}$ $\pm$ 0.56 & 8.15 $\pm$ 4.59  $\pm$ 1.53  &  0.70 $\pm$ 0.21 $\pm$ 0.55 & 4.36 $\pm$ 1.05 $\pm$ 1.63  \\
QCD multijet       &             ---              &            ---                 &  0.13 $\pm$ 0.06 $\pm$ 0.21 & 1.15 $\pm$ 0.39 $\pm$ 0.74  \\
\hline
SM total           & 3.52 $\pm$ 3.35 $\pm$ 0.56 & 8.59 $\pm$ 4.59  $\pm$ 1.53  &  1.58 $\pm$ 0.23 $\pm$ 0.61 & 7.07 $\pm$ 1.30 $\pm$ 1.84  \\
\hline
Observed           &               3            &                5             &             1               & 2     \\\hline
SUSY(380, 1)        & 2.14 $\pm$ 0.08 $\pm$ 0.38 & 2.16 $\pm$ 0.08  $\pm$ 0.39  &  4.10 $\pm$ 0.10 $\pm$ 0.90 & 1.10 $\pm$ 0.05 $\pm$ 0.27 \\
SUSY(240, 40)       & 1.43 $\pm$ 0.19 $\pm$ 0.21 & 0.96 $\pm$ 0.14  $\pm$ 0.14  &  4.35 $\pm$ 0.27 $\pm$ 0.91 & 3.60 $\pm$ 0.25 $\pm$ 0.83 \\
SUSY(180, 60)       & 0.12 $\pm$ 0.04 $\pm$ 0.02 & 0.04 $\pm$ 0.02  $\pm$ 0.01  &  0.73 $\pm$ 0.11 $\pm$ 0.17 & 2.36 $\pm$ 0.17 $\pm$ 0.54 \\
\hline
\end{tabular}
}
\label{tbl:yieldSysSummary}
\end{table}
There is no evidence for an excess of events with respect to the predicted SM values in any of the signal regions. In \bintwo, two events are observed while 7.07 events are expected. The dominant background source is \wjets events. As a cross-check, data and the prediction in the sideband ($200 < \SumMT < 250$\GeV) are studied: 13 events are observed with an expectation of $17.1 \pm 5.0\,(\text{stat+syst})$ events.
This result indicates that the difference between the observed and predicted event yields in \bintwo can be attributed to a downward fluctuation in the data.

\begin{figure}[!htb]
\centering
\includegraphics[width=0.475\textwidth,keepaspectratio=true]{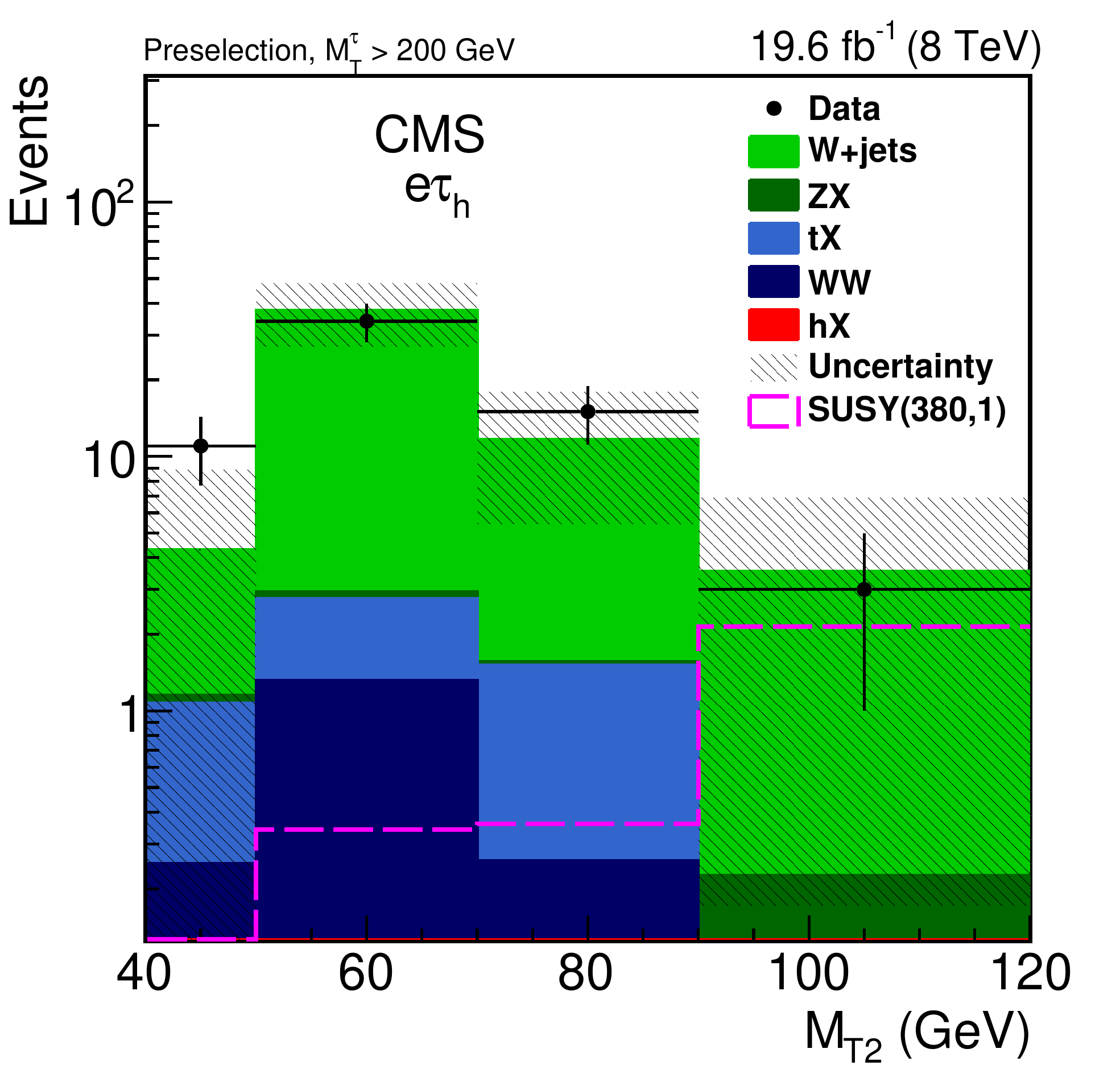}
\includegraphics[width=0.475\textwidth,keepaspectratio=true]{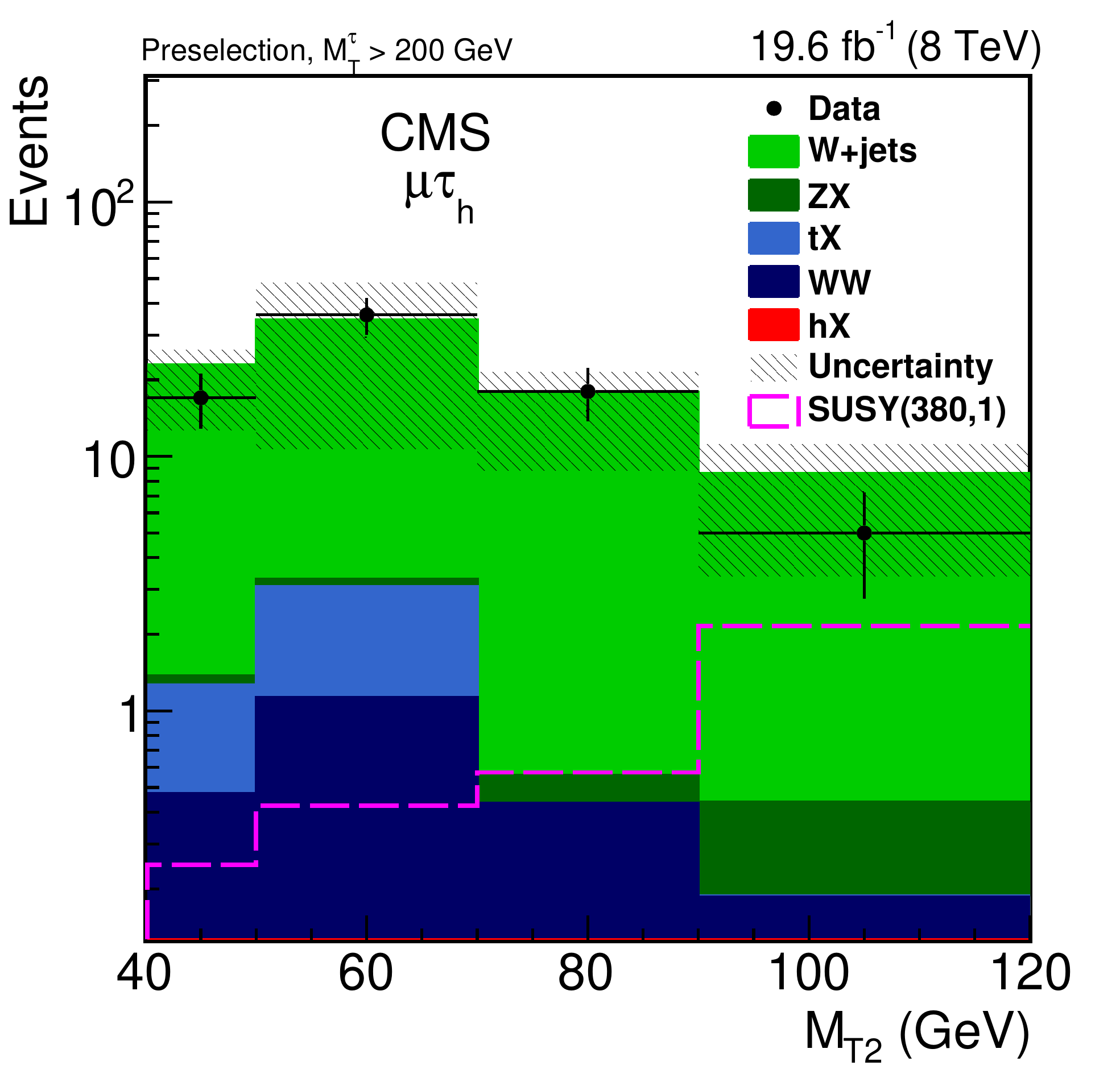}
\includegraphics[width=0.475\textwidth,keepaspectratio=true]{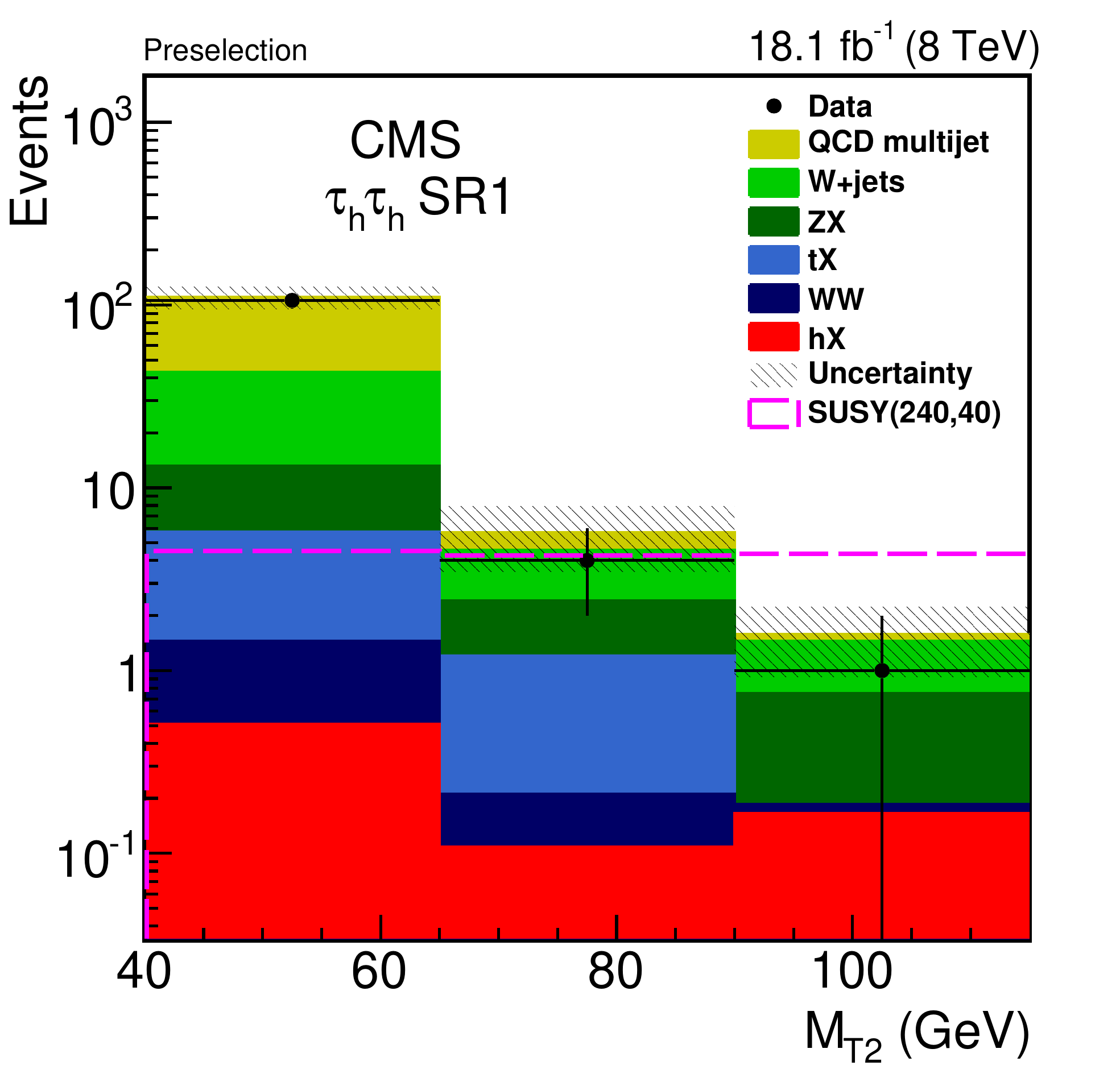}
\includegraphics[width=0.475\textwidth,keepaspectratio=true]{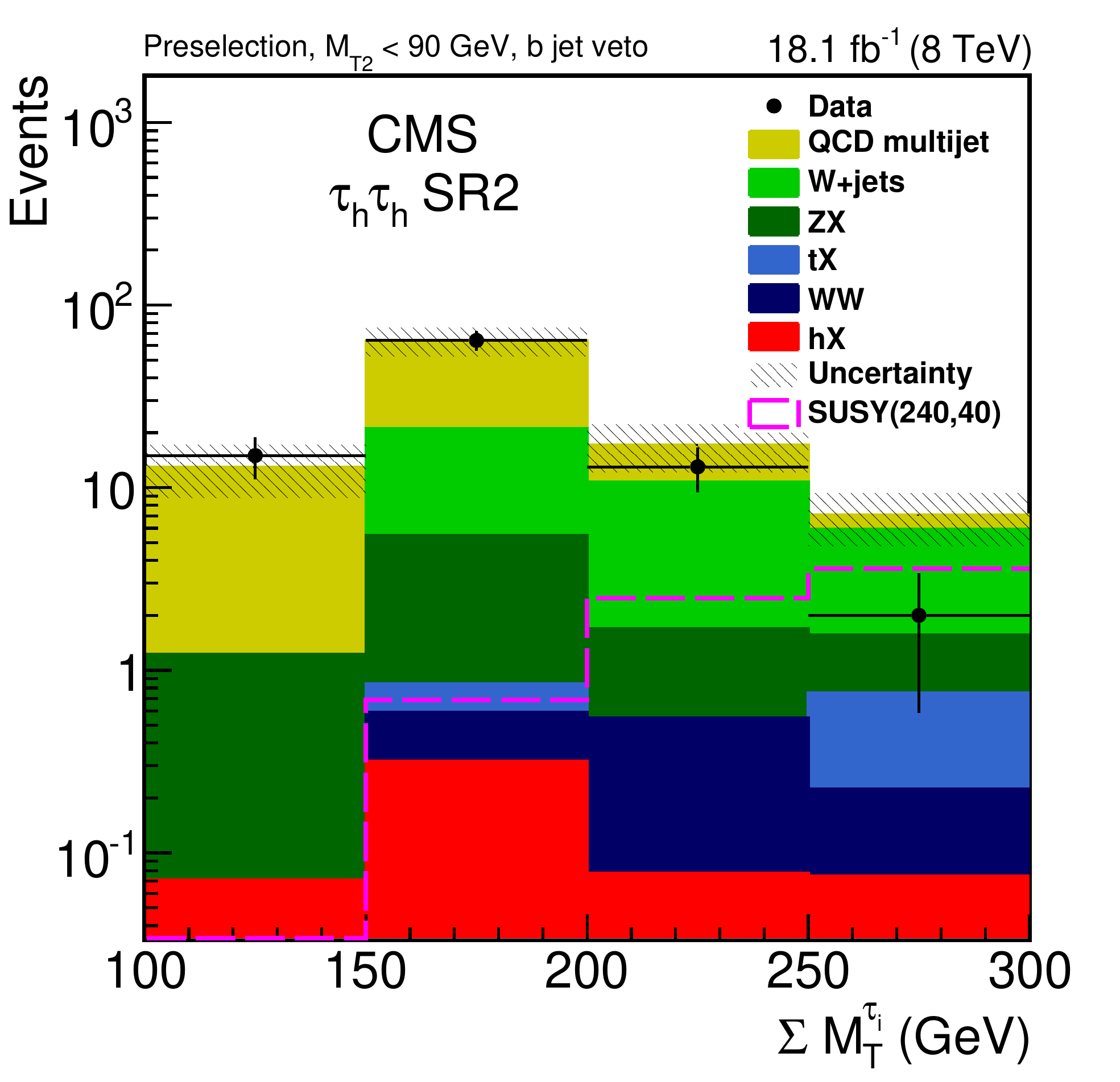}
\caption{The data yield is compared with the SM expectation. In different signal regions,
when a background estimate from data is available, it is used instead of simulation, as described in the text. The signal distribution for a high $\Delta m$ scenario with ${m}_{\PSGcpmDo} = 380$\GeV and ${m}_{\PSGczDo} = 1$\GeV is compared with the yields of \leptonTau channels while a scenario with lower $\Delta m$ (${m}_{\PSGcpmDo} = 240$\GeV and ${m}_{\PSGczDo} = 40$\GeV) is chosen for the comparison in \tauTau channels. The higher values of \mttwo or \SumMT are included in the last bins. The shown uncertainties include the quadratic sum of the statistical and systematic uncertainties.}
\label{fig:yield_final}
\end{figure}

Figure \ref{fig:yield_final}
compares the data and the SM expectation in four search regions. The top row
shows the \mttwo distributions in the \leptonTau channels.
In these plots, the QCD multijet, \wjets, and misidentified lepton contribution from other channels
are based on the estimate described in Section \ref{sect:bkgFake} and labeled as \wjets.
The bottom row shows the \mttwo and \SumMT distributions in the two different signal regions of the \tauTau channel.
The QCD multijet contribution in these plots is obtained using control samples in data, as described in
Section \ref{sect:bkgQCD}. The \wjets contribution in
the last bin of the bottom plots is described in Section \ref{sect:bkgW}, while the contribution to other bins is based on simulated events.
The uncertainty band in these four plots includes both the statistical and systematic uncertainties.

There is no excess of events over the SM expectation.  These results are interpreted in the context
of a simplified model of chargino pair production and decay, which is described in Section~\ref{sect:MCSamples} and corresponds
to the left diagram in Fig.~\ref{fig:Productions}.

A modified frequentist approach, known as the LHC-style CL$_\mathrm{s}$ criterion \cite{read:CLs,Junk:1999kv,ATLAS:2011tau}, is used to
set limits on cross sections at a 95\% confidence level (CL).
The results on the excluded regions are shown in Fig.~\ref{fig:limit_final}.
\begin{linenomath}
\begin{figure}[!htb]
\centering
\includegraphics[width=0.7\textwidth,keepaspectratio=true]{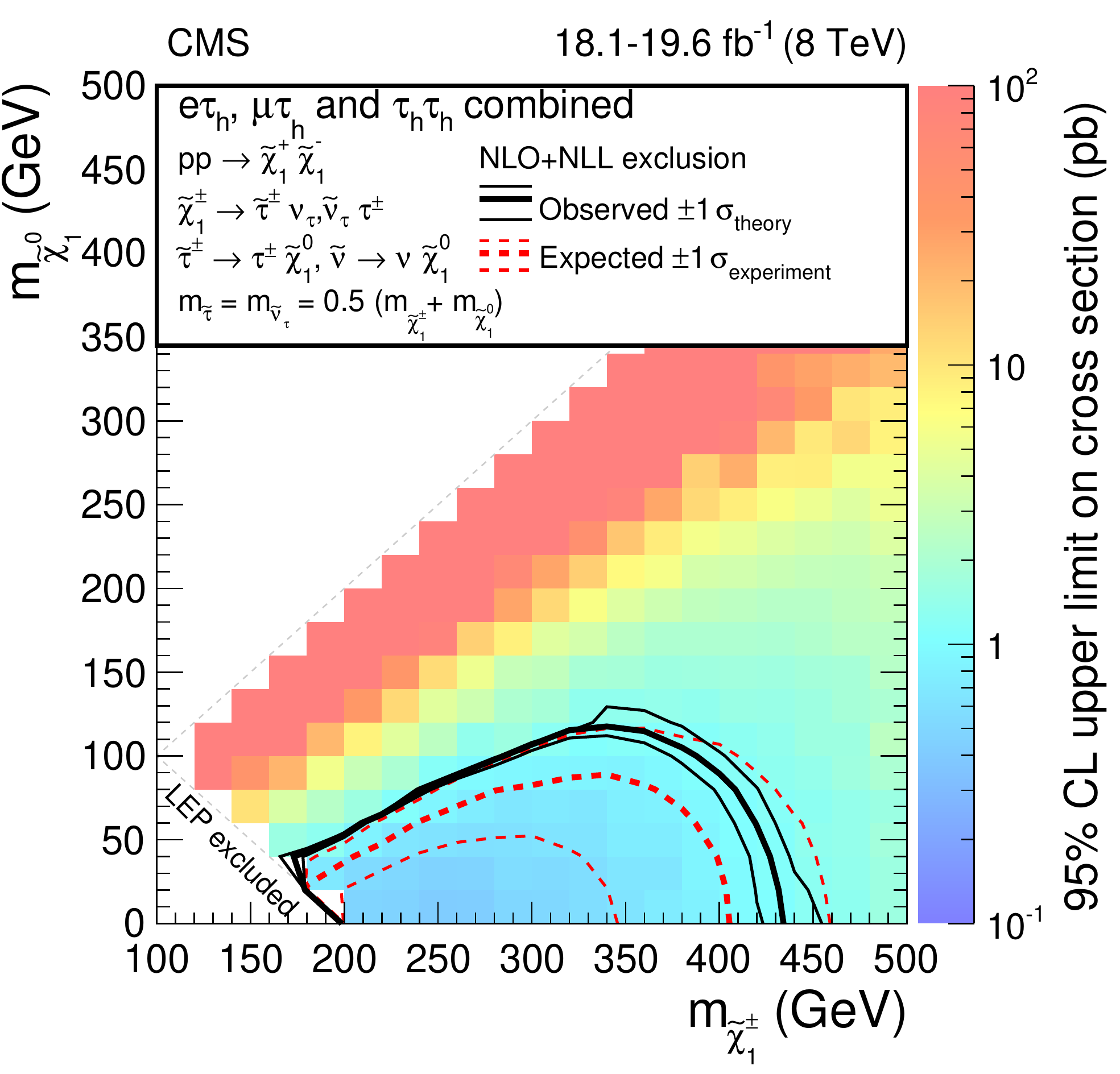}
\caption{Expected and observed exclusion regions in terms of simplified models of
chargino pair production with the total data set of 2012.
The triangle in the bottom-left corner corresponds to  \PSGt masses below 96\GeV, which has been excluded by the LEP experiments~\cite{lepsusy}.
The expected limits and the contours corresponding to $\pm$1 standard deviation from experimental uncertainties are shown as red lines.
The observed limits are shown with a black solid line, while the $\pm$1 standard deviation based on the signal cross section uncertainties
are shown with narrower black lines.}
\label{fig:limit_final}
\end{figure}
\end{linenomath}
Combining all four signal regions,
the observed limits rule out \PSGcpmDo  masses up to  420\GeV  for a massless \PSGczDo.
This can be compared to the ATLAS limit of 345\GeV for a massless \PSGczDo \cite{Aad:2014yka}.
It should be noted that the ATLAS results are based on the \tauTau channel alone. Figure
\ref{fig:limit_tauTau}
\begin{linenomath}
\begin{figure}[!htb]
\centering
\includegraphics[width=0.7\textwidth,keepaspectratio=true]{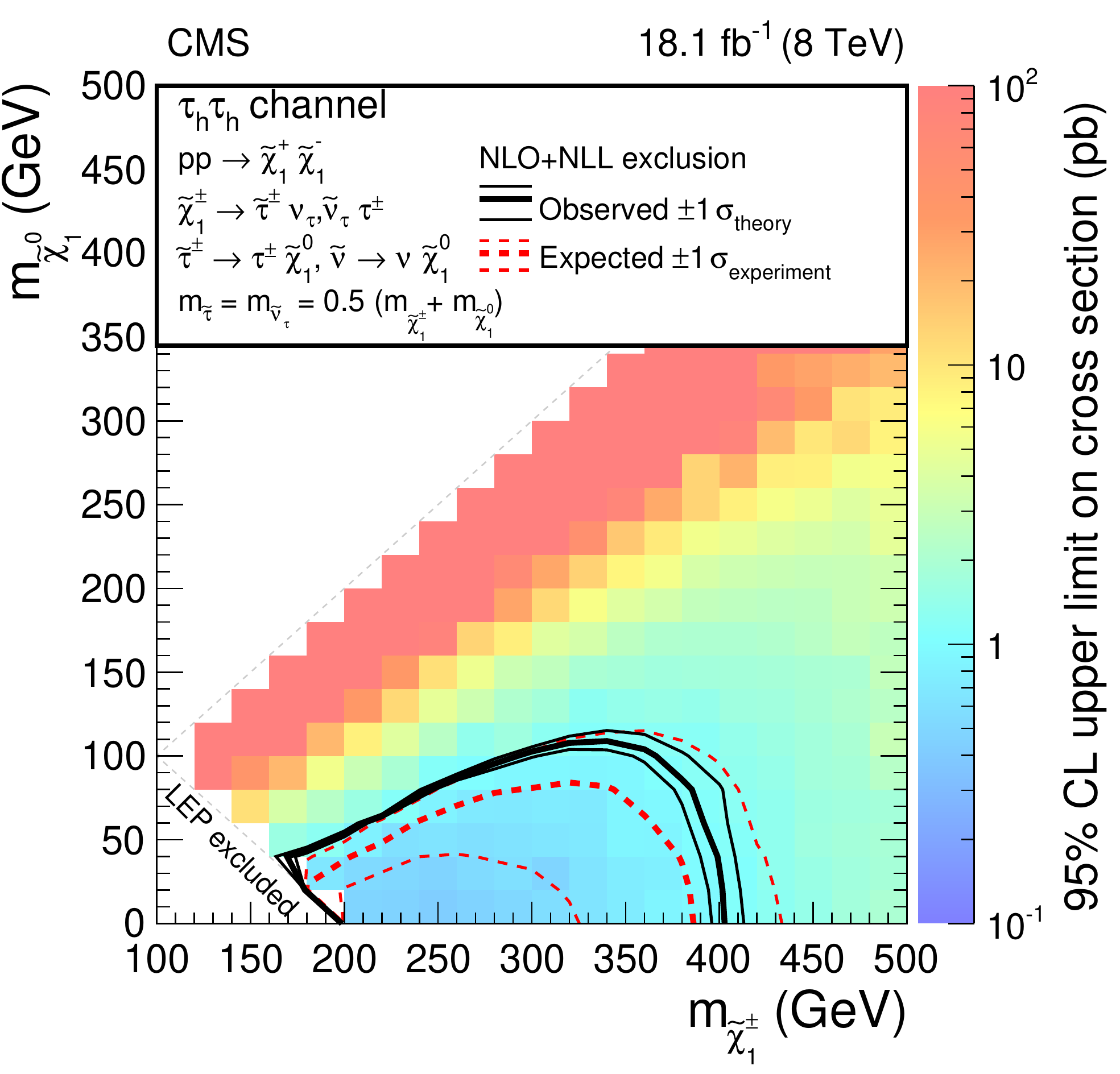}
\caption{Expected and observed exclusion regions in terms of simplified models
in the \tauTau channel. The conventions are the same as Fig.~\ref{fig:limit_final}.}
\label{fig:limit_tauTau}
\end{figure}
\end{linenomath}
shows the results in the \tauTau channel, where the \PSGcpmDo masses are excluded up to 400\GeV for a massless \PSGczDo.
In the whole region, the observed limits are within one standard deviation of the expected limits.

The results are also interpreted to set limits on $\PSGt\PSGt$ production,
which corresponds to the right diagram in Fig.~\ref{fig:Productions}.
In this simplified model, two \PSGt particles are directly produced from the pp  collision and decay promptly to two $\tau$ leptons and two neutralinos.
The effect of the two $\ell\Tau$ channels are found to be negligible and therefore are not considered.
To calculate the production cross section, \PSGt is defined as the left-handed \PSGt gauge eigenstates \cite{Fuks:2013lya}.
Since the cross section for direct production of sleptons is lower, no point is excluded and a 95\% CL upper limit is set on
the cross section  as a function of the \PSGt mass.
Figure \ref{fig:limit_stau_stau} displays the ratio of the
\begin{linenomath}
\begin{figure}[!htb]
\centering
\includegraphics[width=1.0\textwidth,keepaspectratio=true]{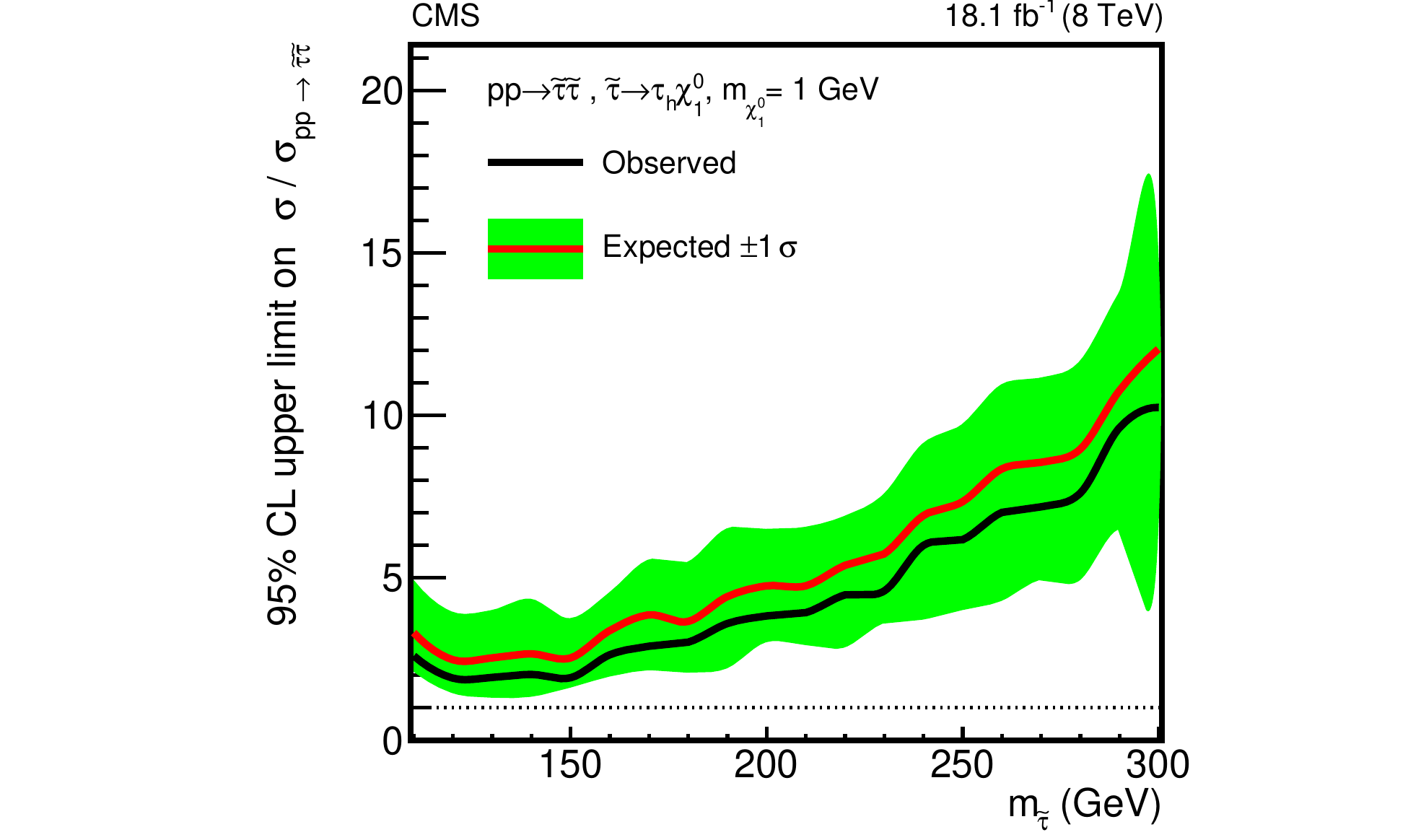}
\caption{Upper limits at 95\% confidence level on the left-handed $\PSGt$ pair production cross section in the \tauTau channel. The mass of \PSGczDo is 1\GeV.
The best observed (expected) upper limit on the cross section is 43\,(56)\unit{fb} for $m_{\PSGt}=150\GeV$
which is almost two  times larger than the theoretical NLO prediction.}
\label{fig:limit_stau_stau}
\end{figure}
\end{linenomath}
obtained upper limit on the cross section and the cross section expected from SUSY (signal strength) versus the mass of the \PSGt particle,
with the \PSGczDo mass set to 1\GeV.
The observed limit is within one standard deviation of  the expected limit.
The best limit, which corresponds to the lowest signal strength, is obtained for $m_{\PSGt}=150\GeV$.
The observed (expected) upper limit on the cross section at this mass is 43 (56)\unit{fb} which is almost two  times larger than the theoretical NLO prediction.

\section{Summary}
\label{sect:conclusion}
A search for SUSY in the $\tau\tau$ final state has been performed where the
$\tau$ pair is produced in a cascade decay from the electroweak production of a chargino pair.  The data analyzed were from pp collisions
at $\sqrt{s} = 8$\TeV collected by the CMS detector at the LHC corresponding to integrated luminosities between 18.1 and 19.6\fbinv.
To maximize the sensitivity, the selection criteria are optimized for \tauTau (small $\Delta m$),
\tauTau (large $\Delta m$), and \leptonTau channels using the variables \mttwo, \tauMT, and \SumMT.
The observed number of events is consistent with the SM expectations.
In the context of simplified models, assuming that the third generation
sleptons are the lightest sleptons and that their masses lie midway between
that of the chargino and the neutralino, charginos lighter than 420\GeV
for a massless neutralino are excluded at a 95\% confidence level.
For neutralino masses up to 100\GeV, chargino masses up to 325\GeV are excluded
at a 95\% confidence level.
Upper limits on the direct $\PSGt\PSGt$ production cross section are also provided,
and the best limit obtained is for the massless neutralino scenario, which is two times
larger than the theoretical NLO cross sections.

\clearpage
\begin{acknowledgments}
\hyphenation{Bundes-ministerium Forschungs-gemeinschaft Forschungs-zentren} We congratulate our colleagues in the CERN accelerator departments for the excellent performance of the LHC and thank the technical and administrative staffs at CERN and at other CMS institutes for their contributions to the success of the CMS effort. In addition, we gratefully acknowledge the computing centres and personnel of the Worldwide LHC Computing Grid for delivering so effectively the computing infrastructure essential to our analyses. Finally, we acknowledge the enduring support for the construction and operation of the LHC and the CMS detector provided by the following funding agencies: the Austrian Federal Ministry of Science, Research and Economy and the Austrian Science Fund; the Belgian Fonds de la Recherche Scientifique, and Fonds voor Wetenschappelijk Onderzoek; the Brazilian Funding Agencies (CNPq, CAPES, FAPERJ, and FAPESP); the Bulgarian Ministry of Education and Science; CERN; the Chinese Academy of Sciences, Ministry of Science and Technology, and National Natural Science Foundation of China; the Colombian Funding Agency (COLCIENCIAS); the Croatian Ministry of Science, Education and Sport, and the Croatian Science Foundation; the Research Promotion Foundation, Cyprus; the Secretariat for Higher Education, Science, Technology and Innovation, Ecuador; the Ministry of Education and Research, Estonian Research Council via IUT23-4 and IUT23-6 and European Regional Development Fund, Estonia; the Academy of Finland, Finnish Ministry of Education and Culture, and Helsinki Institute of Physics; the Institut National de Physique Nucl\'eaire et de Physique des Particules~/~CNRS, and Commissariat \`a l'\'Energie Atomique et aux \'Energies Alternatives~/~CEA, France; the Bundesministerium f\"ur Bildung und Forschung, Deutsche Forschungsgemeinschaft, and Helmholtz-Gemeinschaft Deutscher Forschungszentren, Germany; the General Secretariat for Research and Technology, Greece; the National Scientific Research Foundation, and National Innovation Office, Hungary; the Department of Atomic Energy and the Department of Science and Technology, India; the Institute for Studies in Theoretical Physics and Mathematics, Iran; the Science Foundation, Ireland; the Istituto Nazionale di Fisica Nucleare, Italy; the Ministry of Science, ICT and Future Planning, and National Research Foundation (NRF), Republic of Korea; the Lithuanian Academy of Sciences; the Ministry of Education, and University of Malaya (Malaysia); the Mexican Funding Agencies (BUAP, CINVESTAV, CONACYT, LNS, SEP, and UASLP-FAI); the Ministry of Business, Innovation and Employment, New Zealand; the Pakistan Atomic Energy Commission; the Ministry of Science and Higher Education and the National Science Centre, Poland; the Funda\c{c}\~ao para a Ci\^encia e a Tecnologia, Portugal; JINR, Dubna; the Ministry of Education and Science of the Russian Federation, the Federal Agency of Atomic Energy of the Russian Federation, Russian Academy of Sciences, and the Russian Foundation for Basic Research; the Ministry of Education, Science and Technological Development of Serbia; the Secretar\'{\i}a de Estado de Investigaci\'on, Desarrollo e Innovaci\'on and Programa Consolider-Ingenio 2010, Spain; the Swiss Funding Agencies (ETH Board, ETH Zurich, PSI, SNF, UniZH, Canton Zurich, and SER); the Ministry of Science and Technology, Taipei; the Thailand Center of Excellence in Physics, the Institute for the Promotion of Teaching Science and Technology of Thailand, Special Task Force for Activating Research and the National Science and Technology Development Agency of Thailand; the Scientific and Technical Research Council of Turkey, and Turkish Atomic Energy Authority; the National Academy of Sciences of Ukraine, and State Fund for Fundamental Researches, Ukraine; the Science and Technology Facilities Council, UK; the US Department of Energy, and the US National Science Foundation.

Individuals have received support from the Marie-Curie programme and the European Research Council and EPLANET (European Union); the Leventis Foundation; the A. P. Sloan Foundation; the Alexander von Humboldt Foundation; the Belgian Federal Science Policy Office; the Fonds pour la Formation \`a la Recherche dans l'Industrie et dans l'Agriculture (FRIA-Belgium); the Agentschap voor Innovatie door Wetenschap en Technologie (IWT-Belgium); the Ministry of Education, Youth and Sports (MEYS) of the Czech Republic; the Council of Science and Industrial Research, India; the HOMING PLUS programme of the Foundation for Polish Science, cofinanced from European Union, Regional Development Fund, the Mobility Plus programme of the Ministry of Science and Higher Education, the OPUS programme contract 2014/13/B/ST2/02543 and contract Sonata-bis DEC-2012/07/E/ST2/01406 of the National Science Center (Poland); the Thalis and Aristeia programmes cofinanced by EU-ESF and the Greek NSRF; the National Priorities Research Program by Qatar National Research Fund; the Programa Clar\'in-COFUND del Principado de Asturias; the Rachadapisek Sompot Fund for Postdoctoral Fellowship, Chulalongkorn University and the Chulalongkorn Academic into Its 2nd Century Project Advancement Project (Thailand); and the Welch Foundation, contract C-1845.
\end{acknowledgments}

\bibliography{auto_generated}
\appendix
\section{Additional information for new model testing}
\label{sect:model}
In the previous sections, a simplified SUSY model is used to optimize the selection criteria and interpret the results.
Here, the main efficiencies versus generated values are reported, so that these results can be used in
an approximate manner to examine new models in a MC generator-level study.
The number of the passed signal events and its uncertainty that
can be evaluated by a generator-level study
should be combined statistically with the results in Table \ref{tbl:yieldSysSummary} to find the upper limit
on the number of signal events
and decide if a model is excluded or still allowed according to  the analysis presented in this paper.

Efficiencies are provided as a function of the kinematic properties (\eg, \pt) of visible $\tau$ lepton
decay products at the generator level. The visible $\tau$ lepton (\visTau), if it decays leptonically,
is defined as the 4-vector of the light charged lepton. In hadronic decays, \visTau is the difference
between the 4-vector of the $\tau$ lepton and neutrino in the hadronic decay.
The visible $\tau$ objects are required to pass the offline kinematic selection criteria ($\eta$ and \pt requirements).
The \genMET variable is defined as the magnitude of the negative vector sum of the \visTau pairs in the transverse plane.
The 4-vector of the \visTau objects and \genMET are used to calculate the \mt of the \visTau objects and  also the generator-level \mttwo.
All efficiencies are derived using the SUSY chargino pair production sample.
The chargino mass is varied from 120 to 500\GeV and the neutralino mass from 1 to 500\GeV.
Table \ref{tbl:EffTauLep}
\begin{table}[!htb]
\centering
\topcaption{Efficiencies to select a lepton or \Tau in different channels. Here, $\Tau^1$ and $\Tau^2$ stand for leading and subleading (in \pt) \Tau in the \tauTau channel. Zero for the efficiency shows the region where the generated $\tau$ leptons do not pass the kinematical and geometrical selection cuts.}
\begin{tabular}{cccccc}
\hline
\pt($\ell$  or  $\visTau$) (\GeVns{})         & e for $\eTau$ &  $\mu$ for $\muTau$  & \Tau for $\ell\Tau$    &  $\Tau^1$ for \tauTau & $\Tau^2$ for \tauTau\\
\hline
20--30                     &    0.27       &    0.80              &         0.20           &       0               & 0    \\
30--40                     &    0.68       &    0.86              &         0.36           &       0               & 0    \\
40--60                     &    0.75       &    0.87              &         0.42           &       0.04            & 0.61 \\
60--80                     &    0.80       &    0.89              &         0.47           &       0.14            & 0.69 \\
80--120                    &    0.83       &    0.90              &         0.50           &       0.26            & 0.70 \\
120--160                   &    0.86       &    0.90              &         0.51           &       0.31            & 0.70 \\
160--200                   &    0.87       &    0.91              &         0.51           &       0.34            & 0.71 \\
$>$200                     &    0.89       &    0.92              &         0.51           &       0.37            & 0.71 \\\hline
\end{tabular}
\label{tbl:EffTauLep}
\end{table}
shows the efficiencies for selecting a lepton or \Tau for different channels versus \pt(\visTau).
These efficiencies include the scale factors, and efficiencies of object identification, isolation, and trigger.
Table \ref{tbl:EffMet}
\begin{table}[!htb]
\centering
\topcaption{Efficiencies of the \MPT requirement in all channels versus \genMET.}
\begin{tabular}{cc}
\hline\\[-2.2ex]
\genMET  (\GeVns{})        & All channels\\
\hline
0--10                   &    0.52 \\
10--20                  &    0.58 \\
20--30                  &    0.68 \\
30--40                  &    0.79 \\
40--50                  &    0.87 \\
50--60                  &    0.93 \\
60--70                  &    0.95 \\
70--80                  &    0.97 \\
80--90                  &    0.98 \\
90--100                 &    0.98 \\
100--120                &    0.99 \\
120--140                &    0.99 \\
140--160                &    0.99 \\
$>$160                 &    1.00  \\\hline
\end{tabular}
\label{tbl:EffMet}
\end{table}
shows the efficiencies in all channels to pass the $\MPT > 30$\GeV requirement as a function of the \genMET.
Table \ref{tbl:EffMass}
\begin{table}[!htb]
\centering
\topcaption{Efficiencies of the invariant mass requirements in different channels versus generated mass.}
\begin{tabular}{ccc}
\hline
Generated mass (\GeVns{})  & $\ell\Tau$  &  \tauTau \\
\hline
5--10                 &    0.10     &   0   \\
10--15                &    0.23     &   0.20   \\
15--20                &    0.97     &   0.90   \\
20--25                &    0.99     &   0.94   \\
25--30                &    1.00     &   0.98   \\
30--35                &    0.99     &   1.00   \\
35--40                &    0.98     &   1.00   \\
40--45                &    0.84     &   0.99   \\
45--50                &    0.16     &   0.95   \\
50--55                &    0.04     &   0.68   \\
55--60                &    0.02     &   0.18   \\
60--65                &    0.01     &   0.06   \\
65--70                &    0.04     &   0.03   \\
70--75                &    0.23     &   0.05   \\
75--80                &    0.78     &   0.15   \\
80--85                &    0.91     &   0.40   \\
85--90                &    0.96     &   0.78   \\
90--95                &    0.97     &   0.92   \\
95--100               &    0.98     &   0.95   \\
100--105              &    1.00     &   0.98   \\
105--110              &    1.00     &   0.99   \\
$>$110                &    1.00     &   1.00   \\\hline
\end{tabular}
\label{tbl:EffMass}
\end{table}
shows the efficiencies in different channels to pass the requirement of the reconstructed invariant mass
versus the invariant mass of the
\visTau pair (generated mass). The requirements
on the invariant mass of the reconstructed pair are ($>$15\GeV) and ($<$45 or $>$75\GeV) for the $\ell\Tau$ channels
and ($<$55 or $>$85\GeV) for the \tauTau channel.
The efficiencies of the ($\mttwo >90$\GeV) requirement in $\ell\Tau$ signal region and \tauTau \binone are listed in Table \ref{tbl:EffMT2}.
\begin{table}[!htb]
\centering
\topcaption{Efficiencies of the $\mttwo > 90$\GeV requirement in all channels versus generated \mttwo.}
\begin{tabular}{ccc}
\hline
Generated \mttwo (\GeVns{})    & $\ell\Tau$  &  \tauTau \binone \\
\hline
20--40                    &    0.002    &   0.01  \\
40--50                    &    0.01     &   0.01  \\
50--60                    &    0.02     &   0.03  \\
60--70                    &    0.05     &   0.07  \\
70--80                    &    0.13     &   0.17  \\
80--90                    &    0.35     &   0.44  \\
90--100                   &    0.65     &   0.73  \\
100--110                  &    0.82     &   0.88  \\
110--120                  &    0.90     &   0.94  \\
120--130                  &    0.93     &   0.97  \\
130--140                  &    0.95     &   0.98  \\
140--160                  &    0.96     &   0.98  \\
160--180                  &    0.97     &   0.99  \\
$>$180                    &    0.97     &   1.00  \\\hline
\end{tabular}
\label{tbl:EffMT2}
\end{table}
Table \ref{tbl:EffTauMT}
\begin{table}[!htb]
\centering
\topcaption{Efficiencies of the \tauMT requirement in $\ell\Tau$ channels versus generated \tauMT.}
\begin{tabular}{cc}
\hline
Generated \tauMT (\GeVns{})  & $\ell\Tau$ \\
\hline
100--125                  &   0.01   \\
125--150                  &   0.03   \\
150--170                  &   0.09   \\
170--190                  &   0.26   \\
190--200                  &   0.51   \\
200--210                  &   0.67   \\
210--230                  &   0.82   \\
230--250                  &   0.91   \\
250--275                  &   0.94   \\
275--300                  &   0.97   \\
$>$300                    &   1.00   \\\hline
\end{tabular}
\label{tbl:EffTauMT}
\end{table}
shows the efficiencies in the $\ell\Tau$ channels to pass the $\tauMT > 200$\GeV requirement versus generated \tauMT.

In the \tauTau \bintwo, the reconstructed \mttwo is constrained to lie between 40 and 90\GeV. Table \ref{tbl:EffMT2SR2}
\begin{table}[!htb]
\centering
\topcaption{Efficiencies of the \mttwo requirement in \tauTau \bintwo versus generated \mttwo. Zero for the efficiency shows the region that the generated
\mttwo is much greater than the selection cut.}
\begin{tabular}{cc}
\hline
Generated \mttwo (\GeVns{})  &  \tauTau \bintwo \\
\hline
0--20     & 	0.08  \\
20--40    & 	0.43  \\
40--50    & 	0.75  \\
50--60    & 	0.82  \\
60--70    & 	0.81  \\
70--80    & 	0.72  \\
80--90    & 	0.49  \\
90--100   & 	0.24  \\
100--110  & 	0.11  \\
110--120  & 	0.05  \\
120--130  & 	0.03  \\
130--140  & 	0.02  \\
140--160  & 	0.01  \\
160--180  & 	0.01  \\
$>$180    & 	0  \\\hline
\end{tabular}
\label{tbl:EffMT2SR2}
\end{table}
shows the efficiencies in \tauTau \bintwo to pass the $40 < \mttwo < 90$\GeV requirement versus generated \mttwo.
The last selection in this channel is
the requirement on \SumMT, which is calculated using the 4-vector of the two \visTau and \genMET. Table \ref{tbl:EffSumMT}
\begin{table}[!htb]
\centering
\topcaption{Efficiencies of the \SumMT requirement in \tauTau \bintwo versus the generated \SumMT.}
\begin{tabular}{ccc}
\hline
Generated \SumMT (\GeVns{})  &  \tauTau \bintwo\\
\hline
80--180       &  0.16  \\
180--200      &  0.19  \\
200--210      &  0.25  \\
210--220      &  0.30  \\
220--230      &  0.36  \\
230--240      &  0.43  \\
240--250      &  0.52  \\
250--260      &  0.55  \\
260--270      &  0.61  \\
270--280      &  0.67  \\
280--290      &  0.68  \\
290--300      &  0.73  \\
300--320      &  0.76  \\
320--340      &  0.77  \\
340--360      &  0.80  \\
360--380      &  0.81  \\
380--400      &  0.81  \\
$>$400        &  0.82  \\\hline
\end{tabular}
\label{tbl:EffSumMT}
\end{table}
shows the efficiencies in \tauTau \bintwo to pass the $\SumMT > 250$\GeV requirement versus generated \SumMT.

To take into account the inefficiencies and misidentifications for charge reconstruction of the
objects, identification of the b-tagged jets, identification of the extra leptons and the minimum angle between
the jets and \MET in the transverse plane, the final yields in \leptonTau and \tauTau
channels must be multiplied by 0.8 and 0.7, respectively.

To use these efficiencies, one needs to multiply the values one after another and combine statistically the
final value with the values reported in Table \ref{tbl:yieldSysSummary}  statistically, to decide if a signal point is excluded.
At the generator level, a pair of $\ell\Tau$ or \tauTau is selected, when the \visTau objects pass
the corresponding offline kinematic selection criteria.

The efficiencies are used to reproduce the yields in the SMS plane. The results are in agreement with the yields from the full chain of
simulation and reconstruction within $\sim$30\%.
A user of these efficiencies should be aware that some assumptions can be
broken close to the diagonal (very low mass difference between chargino and neutralino) and these efficiencies cannot be used.
This compressed region requires a separate analysis,
because the mass difference of the parent particle and its decay products is comparable
to the energy threshold used in this analysis to select the objects.

\cleardoublepage \section{The CMS Collaboration \label{app:collab}}\begin{sloppypar}\hyphenpenalty=5000\widowpenalty=500\clubpenalty=5000\textbf{Yerevan Physics Institute,  Yerevan,  Armenia}\\*[0pt]
V.~Khachatryan, A.M.~Sirunyan, A.~Tumasyan
\vskip\cmsinstskip
\textbf{Institut f\"{u}r Hochenergiephysik,  Wien,  Austria}\\*[0pt]
W.~Adam, E.~Asilar, T.~Bergauer, J.~Brandstetter, E.~Brondolin, M.~Dragicevic, J.~Er\"{o}, M.~Flechl, M.~Friedl, R.~Fr\"{u}hwirth\cmsAuthorMark{1}, V.M.~Ghete, C.~Hartl, N.~H\"{o}rmann, J.~Hrubec, M.~Jeitler\cmsAuthorMark{1}, A.~K\"{o}nig, I.~Kr\"{a}tschmer, D.~Liko, T.~Matsushita, I.~Mikulec, D.~Rabady, N.~Rad, B.~Rahbaran, H.~Rohringer, J.~Schieck\cmsAuthorMark{1}, J.~Strauss, W.~Treberer-Treberspurg, W.~Waltenberger, C.-E.~Wulz\cmsAuthorMark{1}
% ...
\vskip\cmsinstskip
% ...
1:~~Also at Vienna University of Technology, Vienna, Austria\\
% ...

\end{sloppypar}
\end{document}